\documentclass[%
 aip,
 amsmath,amssymb,
 reprint,%
author-numerical,%
]{revtex4-1}

\usepackage{graphicx}
\usepackage{dcolumn}
\usepackage{bm}
\usepackage[english]{babel}
\usepackage[utf8]{inputenc}
\usepackage[T1]{fontenc}
\usepackage{diagbox}
\usepackage{braket}
\usepackage{epstopdf}
\usepackage{color,soul}
\usepackage{balance}
\usepackage{amsmath}
\usepackage{amssymb}
\usepackage{autobreak}
\usepackage{subfigure}
\usepackage{graphicx}
\usepackage{float}
\usepackage{tablefootnote}
\usepackage{threeparttable}
\usepackage[section]{placeins}
\usepackage{color}

\begin{document}

\preprint{AIP/123-QED}

\title{Reassessing the potential of TlCl for laser cooling experiments via four-component correlated electronic structure calculations} 



\author{Xiang Yuan}
    \email{xiang.yuan@univ-lille.fr}
    \affiliation{ 
        Université de Lille, CNRS, UMR 8523 - PhLAM - Physique des Lasers, Atomes et Molécules, F-59000 Lille,  France.
    }

    \affiliation{%
        Department of Chemistry and Pharmaceutical Science, Faculty of Science, Vrije Universiteit Amsterdam, de Boelelaan 1083, 1081 HV Amsterdam, The Netherlands.
    }%

\author{Andr\'{e} Severo Pereira Gomes}%
    \email{andre.gomes@univ-lille.fr}
    \affiliation{ 
        Université de Lille, CNRS, UMR 8523 - PhLAM - Physique des Lasers, Atomes et Molécules, F-59000 Lille, France.
    }%



\begin{abstract}
The TlCl molecule has previously been investigated theoretically and proposed as promising candidates for laser cooling searches [X.\ Yuan et.\ al.\ \emph{J.\ Chem.\ Phys.}, \textbf{149}, 094306, 2018]. From these results, the cooling process, which would proceed by transitions between a$^{3}\Pi_{0}^{+}$ and X$^{1}\Sigma_{0}^{+}$ states, had as potential bottleneck the long lifetime (6.04 $\mu s$) of the excited state a$^{3}\Pi_{0}^{+}$, that would prohibit experimentally control the slowing region. Here, we revisit this system by employing four-component Multireference Configuration Interaction (MRCI) calculations, and investigate the effect of such approaches on the calculated transition moments between a$^{3}\Pi_{0}^{+}$ and a$^{3}\Pi_{1}$ excited states of TlCl as well as TlF, the latter serving as a benchmark between theory and experiment. Wherever possible, MRCI results have been cross-validated by, and turned out to be consistent with, four-component equation of motion coupled-cluster (EOM-CC) and polarization propagator (PP) calculations. We find that the results of TlF are very closed to experiment values, while for TlCl the lifetime  of the a$^{3}\Pi_{0}^{+}$ state is now estimated to be 175 ns, which is much shorter than previous calculations indicated, thus yielding a different, more favorable cooling dynamics. By solving the rate-equation numerically, we provide evidence that TlCl could have cooling properties similar to those of TlF. Our investigations also point to the potential benefits of enhancing the stimulated radiation in optical cycle to improve cooling efficiency. 
\end{abstract}

\pacs{}

\maketitle 

\section{\label{sec:level1}Introduction}
 Nowadays, the realization of high precision measurements on atoms and molecules to test violation of time-reversal symmetry, as manifested by an electric dipole moment (EDM) has become a useful tool to search new physics beyond the Standard Model\cite{safronova2018search}. That can be an alternative way of directly searching new particles in collider experiment, which is now estimated as TeV energy scale. A key prerequisite to facilitate such fundamental physics investigations at atomic and molecular level is achieving unprecedented levels of accuracy in high-precision experiments. 
 
 To this end, the use of laser cooling technology provides a powerful method for reducing the noise in the atomic and molecular spectroscopy. 
While widespread for atoms, the cooling of molecules is more challenging; \citeauthor{di2004laser}\cite{di2004laser} had outlined the three conditions for molecular candidates in laser cooling : 1) strong one-photon transition, 2) highly diagonal Franck-Condon factors (FCFs) and 3) no intervening electronic state. Since~\citeauthor{shuman2010laser}\cite{shuman2010laser} first reported the cooling of the SrF molecule, three diatomic molecule (CaF\cite{truppe2017molecules}, YO\cite{hummon20132d}, YbF\cite{lim2018laser}) have been successfully cooled. 

It is interesting to note that, out of these four successfully cooled systems, three contain atoms for which relativistic effects such as spin-orbit coupling play an important role in the resulting molecular electronic structure. Such physical effects yield more complicated energy levels and transition properties that provide additional challenges to the design of cooling schemes, in comparison to species in the upper rows of the periodic table. Therefore, in this context, electronic structure calculations are particularly interesting as a way to prescreen candidate for experiments, due to the fact that highly accurate calculations can be performed for such small systems.

Among species containing heavy elements, Thallim halides (TlX) appeared as an interesting class of systems. TlF is an ideal candidate for measurement of P- and T-violating interactions~\cite{hinds1980experiment,sandars_measurability_1967} because of its high mass and polarizability. Researchers proposed using spin-forbidden transition a$^3\Pi_{1}$-X$^1\Sigma_{0^+}$ to set up cooling optical cycling for TlF \cite{hunter2012prospects}. Many groups had investigated its spectroscopic properties experimentally\cite{clayburn2020measurement, meijer2020lambda}. The wave-number used for optical cycling in TlF is 271.7 nm, reflecting the fact that the corresponding low-lying excited state are rather high, potentially making them less advantageous from an experimental perspective than for species in which such states are lower, as would be the case in heavier TlX species. On the other hand, calculations\cite{Zou2009Comprehensive} on  TlBr, TlI and TlAt have shown that the potential wells of their low-lying excited state deviate too far from ground state, not satisfy condition (2), and thus leaving TlCl as the sole other candidate. 

For TlCl, \citeauthor{yuan2018laser}\cite{yuan2018laser} have carried out a study of its electronic structure  and arrived at the conclusion that a$^3\Pi_{0}^{+}$-X$^1\Sigma_{0^+}$ is an appropriate optical cycling. However, the same work found the radiative lifetime of a$^3\Pi_{0}^{+}$ state to be of about 6 $\mu$s, which is too long for current experiment conditions. This result was however in disagreement with other theoretical works~\cite{li1994relativistic}, which had found the lifetime to be about 800 ns. Interestingly, in these two investigations the final, spin-orbit coupled electronic states have been obtained from scalar relativistic correlated calculations, whose spin-free states are subsequently coupled via spin-orbit configuration interaction (SOCI) calculations. 

Thus, the first objective of this work is to revisit the TlCl system in order to determine what is behind the discrepancies in radiative lifetimes described in the literature, and with that address whether or not it is a system of interest for laser cooling experiments. Given the lack of experimental data on radiative lifetimes for TlCl, we shall also verify the performance of our theoretical approaches with respect to experiment for the TlF system.

As it is found in the literature~\cite{Vallet2000,Weigand:2009dq,Danilo2010,PereiraGomes:2014iy,Kervazo2019} that SOCI calculations can be sensitive to the number of electronic spin-free states entering the SOCI calculation, whether a contracted or uncontracted CI is employed etc., we consider it of interest to attack this problem from a different perspective, with spin-orbit coupling (SOC) interactions being accounted for at mean-field level by using four-component based Hamiltonians, followed by a treatment of electron correlation on a spinor basis, employing the multireference configuration interaction (MRCI) method, accompanied by benchmark calculations with the relativistic equation of motion coupled cluster singles doubles (EOM-CCSD)~\cite{bartlett_coupledcluster_2012} and the Polarization Propagator (PP)~\cite{dreuw_algebraic_2015} approaches, in order to cross-validate the MRCI calculations for ground, excited and transition properties.

This paper is organized as follows. The $ab\;initio$ calculation method is described in section 2. The computational results and the corresponding cooling scheme are presented and discussed in section 3. Finally a brief summary is presented in section 4.

\section{Computational Details}
 The \emph{ab initio} calculations on the electronic states of TlF and TlCl were performed with the DIRAC19 release~\cite{DIRAC19} of the the DIRAC relativistic electronic structure package\cite{saue2020dirac}. In all calculations we employed the four-component Dirac-Coulomb(DC) Hamiltonian, with the usual approximation of the ($SS|SS$) integrals by a Coulombinc correction\cite{visscher1997approximate}. The uncontract Dyall basis sets aae$n$Z \cite{dyall2006relativistic} are selected for Tl atom and the correlation-consistent basis set aug-cc-pV$n$Z \cite{kendall1992electron,woon1993gaussian} for halogen, with $n$ being the basis sets cardinal number ($n = 2, 3, 4$ for double-, triple-, and quadruple-zeta respectively. As a shorthand notation, in the following we shall refer to the different basis sets as $n$Z. 
 
The molecular axis is put along with the z-axis with the center of mass at the original point, and the positive direction is from Tl to X. For  permanent dipole moments (PDM), we use the following bond lengths (\AA), which is the experimental equilibrium distance: 2.0844 (X$^{1}\Sigma_{0}^{+}$, TlF), 2.049 (a$^{3}\Pi_{0}^{+}$, TlF), 2.0745 (a$^{3}\Pi_{1}$, TlF),  2.485 (X$^{1}\Sigma_{0}^{+}$, TlCl), 2.472 (a$^{3}\Pi_{0}^{+}$, TlCl). 2.485 (a$^{3}\Pi_{1}$, TlCl). Since, the PDM of a linear molecule is the first derivative of the energy versus electric field along the molecular z-axis, which form in CBS should be exactly the same as total energy\cite{deng2017permanent}, and therefore we employ the expression :
\begin{equation}
\mu_{CBS}=\mu_{n}-\alpha \exp^{-(n-1)}- \beta \exp^{-(n-1)^2}
\end{equation}
 
We note that for excited states, the complete basis set (CBS) results are extrapolated based on 3Z and 4Z results with the formula~\cite{Kervazo2019,aquilante2020modern}:
\begin{equation}
E_{CBS}(\textbf{R})=\frac{4^3E_{4}(\textbf{R})-3^3E_{3}(\textbf{R})}{4^3-3^3}.
\end{equation}

 
 In this study, we focus on the transitions of a$^{3}\Pi_{1}$-X$^{1}\Sigma_{0}^{+}$ and  a$^{3}\Pi_{0}^{+}$-X$^{1}\Sigma_{0}^{+}$ ($\Omega = 0$ states) in TlF and TlCl. 
 The PDM and the transition dipole moments (TDM) are carried out by multi-reference configuration interaction MRCI method as implemented in the KRCI module\cite{knecht2010large,fleig2003generalized}. In the MRCI calculations the reference configuration space is defined as (8,8) corresponding to 6s 6p 7s orbital of Tl and 2(3)p orbitals of halogen. About $10^{7}$ configurations are included in this work. We note the version of the KRCI module employed in our calculations calculations does not support the use of two-component Hamiltonians.
  
 In addition to MRCI calculations, we have carried out calculations of excitation energy, PDM and TDM by four-component CCSD\cite{visscher1996formulation,shee_analytic_2016}, EOM-CCSD method~\cite{shee_equation--motion_2018} and PP\cite{pernpointner_relativistic_2014,pernpointner_four-component_2018}. As the analytic calculation of expectation values of the excited states is currently not available for both methods in their implementations in DIRAC, we obtained excited state dipole moments through finite field calculations. In these, the component of the dipole moment operator are individually taken as the perturbations with strengths of $\pm$0.0005 a.u, are included at the Hartree-Fock step (corresponding to an orbital-relaxed picture). We also note that for the EOMCC implementation transition moments are also not currently available. In contranst, for PP these are available and will be compared to those obtained with MRCI. In EOM-CCSD and PP, we correlate occupied electrons which are higher -10 a.u, and truncate the virtual space at 20 a.u. 
 
The data, figures and scripts associated to this paper can be obtained as supplemental information at the Zenodo repository~\cite{dataset-lasercooling}.
 

\section{Result and discussion}
\subsection{Permanent dipole moment}

We present the PDMs and vertical excitation energies (T$_{v}$) for MRCI, EOM-CCSD, as well as PP with previous SOCI\cite{liu2020electronic, yuan2018laser} and experiment result\cite{clayburn2020measurement} in Table \ref{tab:table_pdm}. Our results point to an asymptotic convergence of the MRCI PDMs as a function of basis set level for all states under consideration, which leads to a decrease in magnitude of dipole moments for all states of TlF and TlCl at CBS level compared to the smaller basis set results. 

In addition to that, we see that in magnitude the excited state dipole moments are smaller than those for the ground states, and that PDMs for TlF are all smaller in magnitude than those for TlCl. As it turns out, the magnitude of the PDM is larger for TlCl than for TlF in the respective ground states, while the reverse is true for the excited states. We note that all of our results show a dipole with a negative sign, so that a decrease in magnitude actually corresponds to a small build-up of electron density around the Tl atom as the quality of the basis set is improved.

The MRCI results are consistent with the coupled-cluster ones (due to constraints in computational resources, we were not able to carry out EOM calculations with 4Z bases, and thus only present results for 2Z and 3Z); the differences in PDMs between the two approaches is often, in absolute value, of around 0.3 D and no larger than 0.5 D for all states considered. It is interesting to note that differences between the methods tend to be smaller for 3Z than for 2Z bases, with the coupled cluster results varying less than MRCI ones when going from 2Z to 3Z, and as such we expect that our 3Z results can serve as a semi-quantitative comparison. This gives us confidence that our 4Z and CBS MRCI results should be reliable.

For the coupled cluster ground states, for which we can also calculate PDMs analytically, we observe that the finite-field and analytic derivative results are very close to each other (differences around 0.03-0.04 D), providing indications that (a) orbital relaxation is not particularly important for such species and (b) the finite-field results for the different electronic states are reliable.

Finally, compared to the SOCI results of \citeauthor{yuan2018laser}\cite{yuan2018laser}, we see that for the ground state, SOCI and our MRCI results match rather well. For excited states, the situation is different, and for some states such as the $^3\Pi_1$, we observe significant differences between methods for both TlF and TlCl. Furthermore, for TlF there are also important differences for the $^3\Pi_0^+$ state. 

In comparison to experiment, the measurements of the PDM of the a$^{3}\Pi_{1}$ state of TlF by~\citeauthor{clayburn2020measurement}\cite{clayburn2020measurement} yielding a value of -2.28(7) D, are in very good agreement with our MRCI 4Z (-2.46 D) or CBS (-2.47 D) results, or our EOM-CCSD 3Z results (-2.47 D). From that, and the very close matching of our MRCI and CCSD results for TlCl, we expect that our results should provide a good estimate for the experimental value. We believe future high-resolution PDM measurements for more states would be highly desirable as a test, and possible confirmation, of our results.

Concerning excitation energies (T$_{v}$), for TlF we see significant variations with basis set size for MRCI, yield decreasing excitation energies as basis sets quality is improved--roughly a 2000 cm$^{-1}$ decrease when passing from 2Z to 3Z for both $\Pi$ states, and around 1000 cm$^{-1}$ when passing from 3Z to 4Z, and another 1000 cm$^{-1}$ when passing from 4Z to CBS, adding up to changes of nearly 5000 cm$^{-1}$ from 2Z to CBS. This trend is qualitative the same for EOM calculations, though the changes are much smaller (around 300-400 cm$^{-1}$ from 2Z to 3Z). 

For TlCl, on the other hand, we observe much smaller variations with changing basis sets for MRCI (less than 1000 cm$^{-1}$ between 2Z to CBS results), with excitation energies increasing as basis sets are improved. As for TlF, EOM trends follow the MRCI ones, and energy changes are again much less importan than those for MRCI. 

In contrast to MRCI and EOM, PP results for all excited states show an increase in excitation energies with increased basis set size, with changes between 2Z and 3Z basis being of about 1000 cm$^{-1}$ in all cases. It is interesting to note that the 3Z PP excitation energies are generally lower but not far from MRCI CBS results for both molecules. While that seems fortuitous for TlF, due to the large variations in excitation energies for MRCI, for TlCl it appears that all three correlated approaches do indeed show quite similar performances. 

We see that for both species, the T$_{v}$ EOM results are very much in line with the experimental T$_{e}$ values (given the nature of the excited states, the calculated T$_{v}$ values should in effect be quite close to the T$_{e}$ ones), meaning that MRCI energies can reliably reflect experimental excitation energies.

The difference between SOCI and the current excitation energies for TlF is striking, with SOCI overestimating the four-component results by 2000 to 3000 cm$^{-1}$, depending on the excited state. Taken together with the (a) strongly underestimated magnitude of the excited state dipole moment with respect to experiment; and (b) the rather good agreement for calculated grond-state dipole moments, it would appear that the SOCI calculations of~\citeauthor{yuan2018laser}\cite{yuan2018laser} are somewhat unbalaced in their description of excited states, with respect to the ground-state. Interestingly, for TlCl the SOCI calculations match rather well both the four-component excited state energies, and ground and excited state dipole moments. This suggests that any issues with SOCI calculations would not be so much in the description of the different electronic states of TlCl, but rather in on the transition properties, which we turn our attention to next.



\begin{table*}[]
\begin{threeparttable}
\renewcommand\arraystretch{1}
\footnotesize
\caption{\label{tab:table_pdm}
The computed permanent dipole moments (D) and vertical excitation energies (T$_{v}$)(cm$^{-1}$) for the different states under consideration for TlF and TlCl. }
\begin{ruledtabular}
\begin{tabular}{ccccccccccccccc}
        &  & & \multicolumn{2}{c}{2Z} &  \multicolumn{2}{c}{3Z} & \multicolumn{2}{c}{4Z} & \multicolumn{2}{c}{CBS} & \multicolumn{2}{c}{SOCI\cite{yuan2018laser,liu2020electronic}}  & \multicolumn{2}{c}{Exp}\\
\cline{4-5}
\cline{6-7}
\cline{8-9}
\cline{10-11}
\cline{12-13}
\cline{14-15}
Molecule & State & Method 
& \scriptsize{PDM} & \scriptsize{T$_{v}$}    
& \scriptsize{PDM} & \scriptsize{T$_{v}$}    
& \scriptsize{PDM} & \scriptsize{T$_{v}$}   
& \scriptsize{PDM} & \scriptsize{T$_{v}$} 
& \scriptsize{PDM} & \scriptsize{T$_{v}$} 
& \scriptsize{PDM} & \scriptsize{T$_{v}$\tnote{c}}  \\
\hline
TlF      & X$^{1}\Sigma_{0}^{+}$   & MRCI   & -4.16  &    0      & -3.88  &    0      & -3.79  &    0      & -3.74   &0 & -3.67    &   0      &       \\
         &       & CCSD\tnote{a}    &  -4.37      & 0         & -4.32  &      0    &        &          &         &&          &         &       \\
         &       & CCSD\tnote{b}    & -4.33  & 0        & -4.29  &       0   &        &          &         &&          &         &       \\
&&&&&&&&&&&&&\\
         & a$^{3}\Pi_{0}^{+}$    & MRCI   & -3.15  & 36825& -2.81  & 34695 & -2.76  & 33708& -2.74   &32990& -1.46    & 37025   &   &35164    \\
         &       & EOM\tnote{a}    &    -2.67    &   34790       & -2.69  & 35082 &        &          &         &&          &         &       \\
         &       & PP\tnote{a}    &        &   31592       &   &  32414 &        &          &         &&          &         &       \\
&&&&&&&&&&&&&\\
& a$^{3}\Pi_{1}$    & MRCI   & -2.90   & 40070 & -2.47  & 37921 & -2.46  & 36507 & -2.47   &35474& -1.26    & 38535   & -2.28 &36864\\
         &       & EOM\tnote{a}    &   -2.45     &  36475        & -2.47  & 36782 &        &          &         &&          &         &       \\
         &       & PP\tnote{a}    &        &   32920       &   & 33719 &        &          &         &&          &         &       \\
&&&&&&&&&&&&&\\
TlCl     & X$^{1}\Sigma_{0}^{+}$   & MRCI   & -4.60   &   0       & -4.46  &   0       & -4.42  &    0      & -4.4    &0& -4.32    &  0       &       \\
         &       & CCSD\tnote{a}    &  -4.66      &   0       & -4.65  &   0       &        &          &         &&          &         &       \\
         &       & CCSD\tnote{b}    &   -4.63     &  0        & -4.64  &   0       &        &          &         &&          &         &       \\
&&&&&&&&&&&&&\\
& a$^{3}\Pi_{0}^{+}$    & MRCI   & -2.43  & 31130    & -2.19  & 31630    & -2.13  & 31813    & -2.1    & 31947& -2.08    &  31438       &    &31054   \\
         &       & EOM\tnote{a}    &   -1.78     &  31095        & -1.80   & 31193 &        &          &         &          &&         &       \\
          &       & PP\tnote{a}    &        &  30801        &    &  31338 &        &          &         &          &&         &       \\
&&&&&&&&&&&&&\\
         & a$^{3}\Pi_{1}$\tnote{*}    & MRCI   & -1.83  & 33366         & -1.51  &  34594        & -1.48  &    34711      & -1.47   &34797  & -1.74    & 32526        &       \\
         &       & EOM\tnote{a}    &   -1.45     &   32369       & -1.29  &  32426        &        &          &         &          &&         &      
\end{tabular}
\begin{tablenotes}
\item [a] Finite filed calculations, equivalent to an orbital-relaxed formulation.
\item [b] Analytic gradient calculation, employing an orbital-unrelaxed formulation~\cite{shee_analytic_2016}.
\item [c] Refer to adiabatic excitation energy value (T$_{e}$) in experiment.
\item [*] a$^{3}\Pi_{1}$ state of TlCl is not a bound state
\end{tablenotes}
\end{ruledtabular}
\end{threeparttable}
\end{table*}

\subsection{Transition properties}

Our results for TDMs, obtained at bond lengths corresponding to the experimental equilibrium distance of the corresponding excited states, are found in table~\ref{tab:table_tdm}. Unlike the excited state properties, for TDMs it is not possible to estimate a CBS value and as such we shall focus on TDM results obtained at 4Z level for MRCI. As EOM TDMs are currently not available, we decided to present TDM values obtained from 4Z PP calculations; from the results discussed previously we expect PP calculations provide a cross-validation of MRCI results, closely matching the MRCI ones for TlCl, but less so for TlF.

It turns out that MRCI and PP results are, in effect, quite close to each other for the different transitions in TlF, differing by less than 0.1 D for the transitions from the ground to each of the $\Pi$ states, the transition to the $^3\Pi_0$ state the PP TDM being larger than the MRCI one, whereas the reverse is true for the transition from the ground to the $^3\Pi_1$. 

In the TlCl case the difference between MRCI and PP is slightly larger than 0.1 D for the transition from ground to the $^3\Pi_0$ state, with the PP value being larger than the MRCI one as in the TlF case. The good agreement between the two four-component approaches for both molecules (discrepancies between MRCI and PP calculations are smaller than 25\%)  gives us confidence in the ability of MRCI of obtaining sufficiently accurate TDMs.

Comparing our current results for TlCl to those in the literature, we observe first that for the transition from ground to the $^3\Pi_1$ state, our results differ by a little over 0.1 D from those from~\citeauthor{yuan2018laser} \cite{liu2020electronic}. Second, we see that the SOCI TDMs of a$^{3}\Pi_{0}^{+}$ are indeed strongly underestimated, differing from ours by nearly 0.7 D. 

The TDM of a$^{3}\Pi_{0}^{+}$ and a$^{3}\Pi_{1}$ of TlCl are slightly larger than the counterparts of TlF for both MRCI and PP, something which is consistent with our understanding that TlCl should have somewhat stronger spin-orbit coupling effects than TlF, whereby further weakening the selection rules making the spin forbidden transition a$^{3}\Pi$-X$^{1}\Sigma^{+}$ in comparison to TlF.

If a comparison of calculated TDMs is already instructive, a more direct comparison to experiment is done through a comparison of lifetimes, also shown in Table \ref{tab:table_tdm}. From the TDMs, we evaluate the Einstein coefficient and lifetime, which had been used in the previous work\cite{yuan2018laser}.
\begin{equation}
 A_{v'v''}=2.142\times10^{10}\times TDM^{2}\times q_{v'v''}\times\Delta E^{3}
 \end{equation}
(where energy difference $\Delta E$, TDM in unit a.u. and $A_{v'v''}$ in unit s$^{-1}$).
and the radiative lifetime is obtained using 
 \begin{equation}
 \tau_{v'}=\frac{1}{\sum_{v''} A_{v'v''}}
 \end{equation}

The vibrational energy level and the corresponding Franck-Condon factors ($q_{v'v''}$) are taken from available experiment\cite{tiemann1988spectroscopic,hunter2012prospects}(a$^{3}\Pi_{1}$ state of TlF) and previous calculations\cite{liu2020electronic,yuan2018laser}(a$^{3}\Pi_{0}^{+}$ state of both TlF and TlCl). The detailed Einstein coefficients $A_{v'v''}$ and vibrational branching $R_{v'v''}$ of transitions are listed in supplement material. 

Given that SOCI PDMs agree rather well with four-component ones for the ground states of TlF and TlCl, we consider this approach (combining ground-state vibrational wavefunctions obtained from SOCI potential energy curves and four-component TDMs at MRCI or PP level) to be reliable.


For TlF, the computed lifetime of the a$^{3}\Pi_{1}$ state is 91 and 153 ns for MRCI and PP, respectively and the former is closer to the experiment value 99(9) ns. For TlCl, the previously calculated 6.04 $\mu$s lifetime of the a$^{3}\Pi_{0}^{+}$ state would correspond to  a huge challenge under current cooling experiment condition. However, we see that on the basis of the current four-component calculations, lifetimes for the  a$^{3}\Pi_{0}^{+}$ state would correspond to 175 ns (MRCI) and 128 ns (PP), respectively, which are both much shorter than previous value and, therefore, may make TlCl more favorable for experimental realization than previously thought.

Given the small deviation between theoretical and experimental lifetimes for the a$^{3}\Pi_{1}$ state of TlF obtained with MRCI, and the systematic agreement between four-component approaches for TDMs, we thus consider the lifetime of a$^{3}\Pi_{0}^{+}$ state of TlCl 175 ns to be a more accurate estimate than the previous estimation by \citeauthor{yuan2018laser}, and shall use the MRCI lifetime in the following assessment of a proposed cooling scheme.

\begin{table}[ht]
\renewcommand\arraystretch{1}
\footnotesize
\caption{\label{tab:table_tdm}
The computed transition dipole moment at $R_{e}\;$ and the corresponding lifetimes}
\begin{ruledtabular}
\begin{tabular}{cccccc}
TlF&Transition&TDM(D)&lifetime(ns)&Reference \\
\hline
&a$^3\Pi_{1}$-X$^1\Sigma_{0}^{+}$&0.837&91 &MRCI\\
&a$^3\Pi_{1}$-X$^1\Sigma_{0}^{+}$&0.673&153 &PP\\
&a$^3\Pi_{1}$-X$^1\Sigma_{0}^{+}$&&99(9)&Exp\cite{hunter2012prospects}\\
&a$^3\Pi_{1}$-a$^3\Pi_{0^+}$&0.114&&MRCI\\
&a$^3\Pi_{1}$-a$^3\Pi_{0^-}$&0.072&&MRCI\\
&a$^3\Pi_{0^+}$-X$^1\Sigma_{0}^{+}$&0.518&278&MRCI\\
&a$^3\Pi_{0^+}$-X$^1\Sigma_{0}^{+}$&0.651&176&PP\\
\hline
TlCl&&& \\
\hline
&a$^3\Pi_{0^+}$-X$^1\Sigma_{0}^{+}$&0.767&175&MRCI\\
&a$^3\Pi_{0^+}$-X$^1\Sigma_{0}^{+}$&0.896&128&PP\\
&a$^3\Pi_{0^+}$-X$^1\Sigma_{0}^{+}$&0.130&6040&Ref\cite{yuan2018laser}\\
&a$^3\Pi_{0^+}$-X$^1\Sigma_{0}^{+}$&&808&Ref\cite{li1994relativistic}\\
&a$^3\Pi_{1}$-X$^1\Sigma_{0}^{+}$&0.946&&MRCI\\
&a$^3\Pi_{1}$-X$^1\Sigma_{0}^{+}$&0.800&&Ref\cite{yuan2018laser}\\
\end{tabular}
\end{ruledtabular}
\end{table}

\subsection{Simulation of laser cooling}

\begin{figure} [htbp]
\centering
\includegraphics [width=\linewidth] {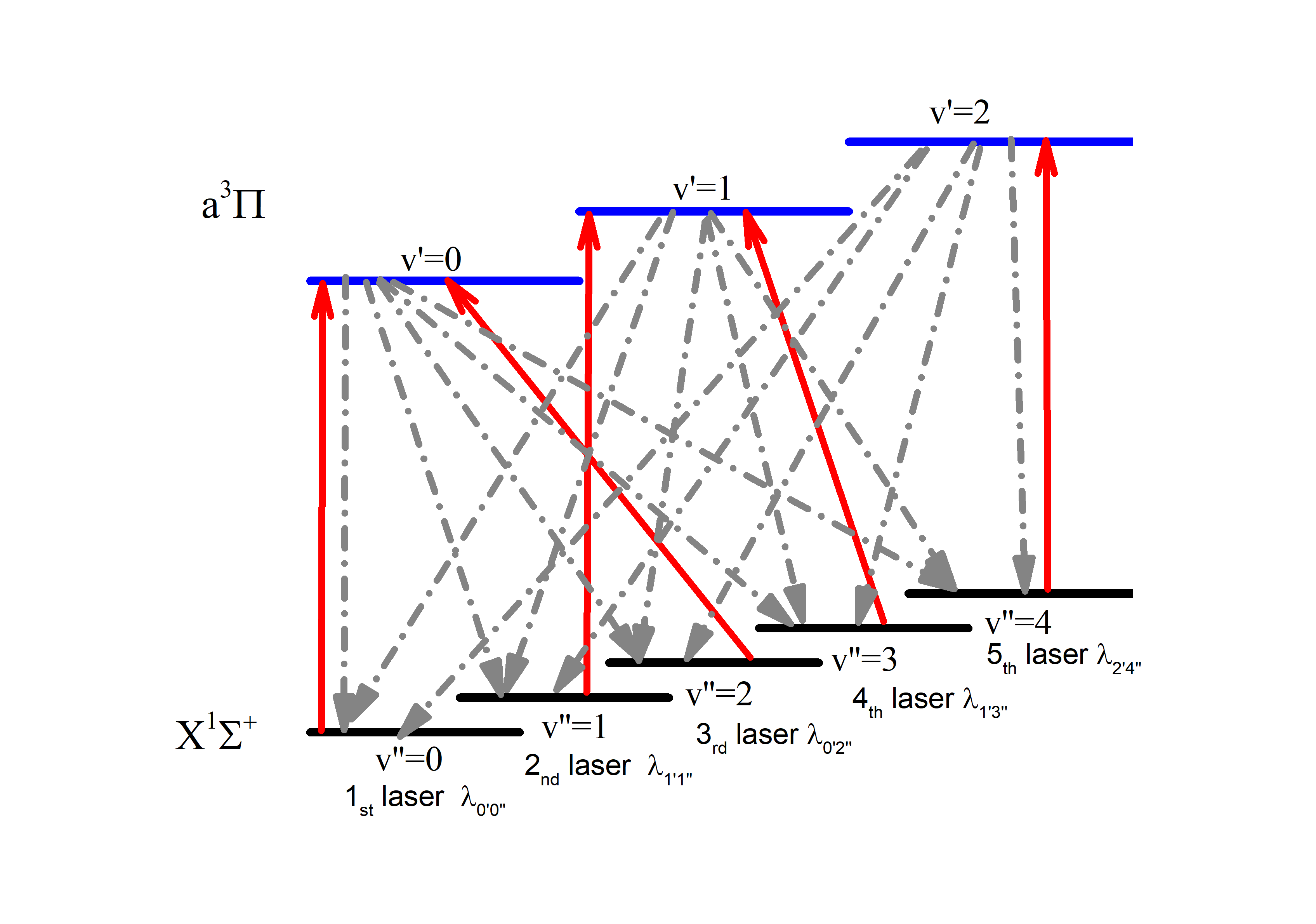}
\caption{\label{fig:epsart_cooling_scheme} The proposed cooling scheme for TlF and TlCl. The excited states are a$^{3}\Pi_{1}$ and a$^{3}\Pi_{0}^{+}$ for TlF and TlCl, respectively. The dashed gray lines are spontaneous decay and the solid red lines are laser-driven transition. }
\end{figure}

Regarding the cooling efficiency and the corresponding length of slowing region is depended on the lifetime, the new value of a$^{3}\Pi_{0}^{+}$ state of TlCl shows a different cooling dynamics and lowers technology difficulty in experiment, compared to the results of \citeauthor{yuan2018laser}\cite{yuan2018laser}. In spite of the changes in TDM the optical cycling scheme for TlF and TlCl, shown in Fig \ref{fig:epsart_cooling_scheme}, still closely follows that originally proposed by~\citeauthor{yuan2018laser}\cite{yuan2018laser}, which is outlined below.

The main pump laser is set at a$^{3}\Pi$(v'=0)- X$^1\Sigma_{0}^{+}$(v''=0) transition with a wavelength $\lambda_{0'0''}$: 272 nm (TlF) and 319 nm (TlCl).  There are four additional lasers for repumping the population of vibrational excited states. For the sake of clarity, we call those lasers in the following order: $\lambda_{0'0''}$ is the first laser, $\lambda_{1'1''}$ is the second lasers,  $\lambda_{0'2''}$ is the third laser, $\lambda_{1'3''}$ is the fourth laser, and $\lambda_{2'4''}$ is the fifth laser. All the wavelength of the lasers are listed in Table \ref{tab:table3}:

\begin{table}[ht]
\renewcommand\arraystretch{1}
\footnotesize
\caption{\label{tab:table3}
The wavelength (nm) of lasers used in cooling process represented by figure~\ref{fig:epsart_cooling_scheme}.}
\begin{ruledtabular}
\begin{tabular}{ccc}
laser&TlF&TlCl\\
\hline
1$_{st}$: $\lambda_{0'0''}$&272&319\\
2$_{rd}$: $\lambda_{1'1''}$&273&320\\
3$_{ed}$: $\lambda_{0'2''}$&279&325\\
4$_{th}$: $\lambda_{1'3''}$&280&326\\
5$_{th}$: $\lambda_{2'4''}$&281&327\\
\end{tabular}
\end{ruledtabular}
\end{table}

To discuss the cooling process in more details, we solve a rate-equation to count the number of photons scattered during the cooling process\cite{nguyen2011challenges}:
\begin{equation}
\frac{\mathrm{d} \textbf{P}}{\mathrm{d} t} =\textbf{MP}
\end{equation}
where \textbf{P} is the vector containing N vibrational levels in order of ascending with energy and M is the N$\times$N matrix consisting of various Einstein coefficient. 

Before simulating the population dynamics, it is necessary to see the influence of vibrational decay process within the ground state X$^1\Sigma_{0}^{+}$. Here, we compute the ratio $\frac{A_{0'0''}}{A_{1''0''}}$ of Einstein coefficient between electronic relaxation $A_{0'0''}$(v'=0) $\to$ (v''=0) and vibrational relaxation $A_{1''0''}$ (v''=1) $\to$ (v''=0). The vibrational transition dipole moment (vTDM) matrix elements over vibrational wave functions of X$^1\Sigma_{0}^{+}$ state is computed with the help of Molcas vibrot module\cite{aquilante2020modern}. 
\begin{equation}
vTDM_{1''0''}=\int \phi_{(v''=1)} R \phi_{(v''=0)} dR
\end{equation}
The ratio for TlF and TlCl is 1.8$\times$10$^{7}$ and 3.0$\times$10$^{7}$, respectively similar as the result of SrF\cite{nguyen2011challenges} 2.5$\times$10$^{7}$. Such large ratio means the vibrational relaxation is very weak, and therefore we chose to ignore it in the following simulation model.

Explicitly, the rate-equation has the form:
\begin{equation}
\begin{aligned}
\frac{\mathrm{d} P_{i}}{\mathrm{d} t} = &-\sum_{j=1}^{j=i-1}A_{ij}P_{i}-\sum_{j=1}^{j=i-1}B_{ij}\rho(\omega_{ij}) P_{i}\\
&-\sum_{j=i+1}^{j=N}B_{ij}\rho(\omega_{ij}) P_{i}+\sum_{j=i+1}^{j=N}A_{ji}P_{j}\\
&+\sum_{j=1}^{j=i-1}B_{ji}\rho(\omega_{ji}) P_{j}+\sum_{j=i+1}^{j=N}B_{ji}\rho(\omega_{ji}) P_{j}
\end{aligned}
\end{equation}
Here, $A_{mn}$, $B_{ij}$, $B_{ji}$ are spontaneous emission, stimulated emission and absorption coefficients, respectively. $\rho(\omega_{ij})$ is the spectral energy density at frequency $\omega_{ij}$. 

After numerically solving the equation, the average number of scattered photons are evaluated by multiplying the obtained population in the excited state of optical cycle by its total radiation rate $A_{ij}+B_{ij}$. The stimulated coefficients $B$ are relative to the $A$ with 
\begin{equation}
B_{ij}=B_{ji}=\frac{\pi^{2}c^{3}}{h\omega_{ij}^{3}}A_{ij}
\end{equation}
where, $h$ is the Planck constant and $c$ is the speed of light. 

\begin{figure}  [htb]
\centering
\subfigure [TlF]{
\label{fig:a}
\begin{minipage}[b]{0.47\textwidth}
\includegraphics [width=\linewidth] {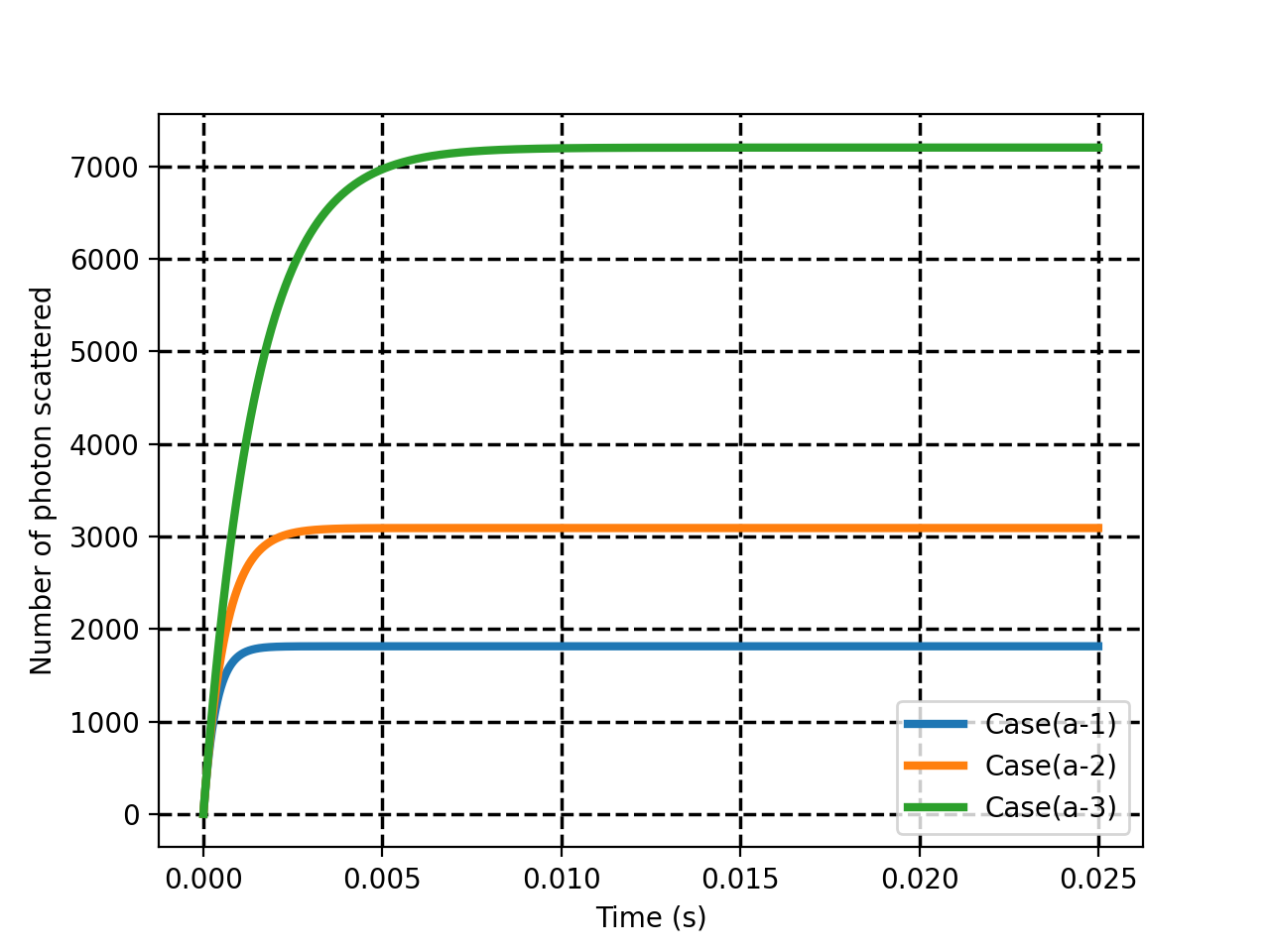}
\end{minipage}
}

\subfigure [TlCl]{
\label{fig:b}
\begin{minipage}[b]{0.47\textwidth}
\includegraphics [width=\linewidth] {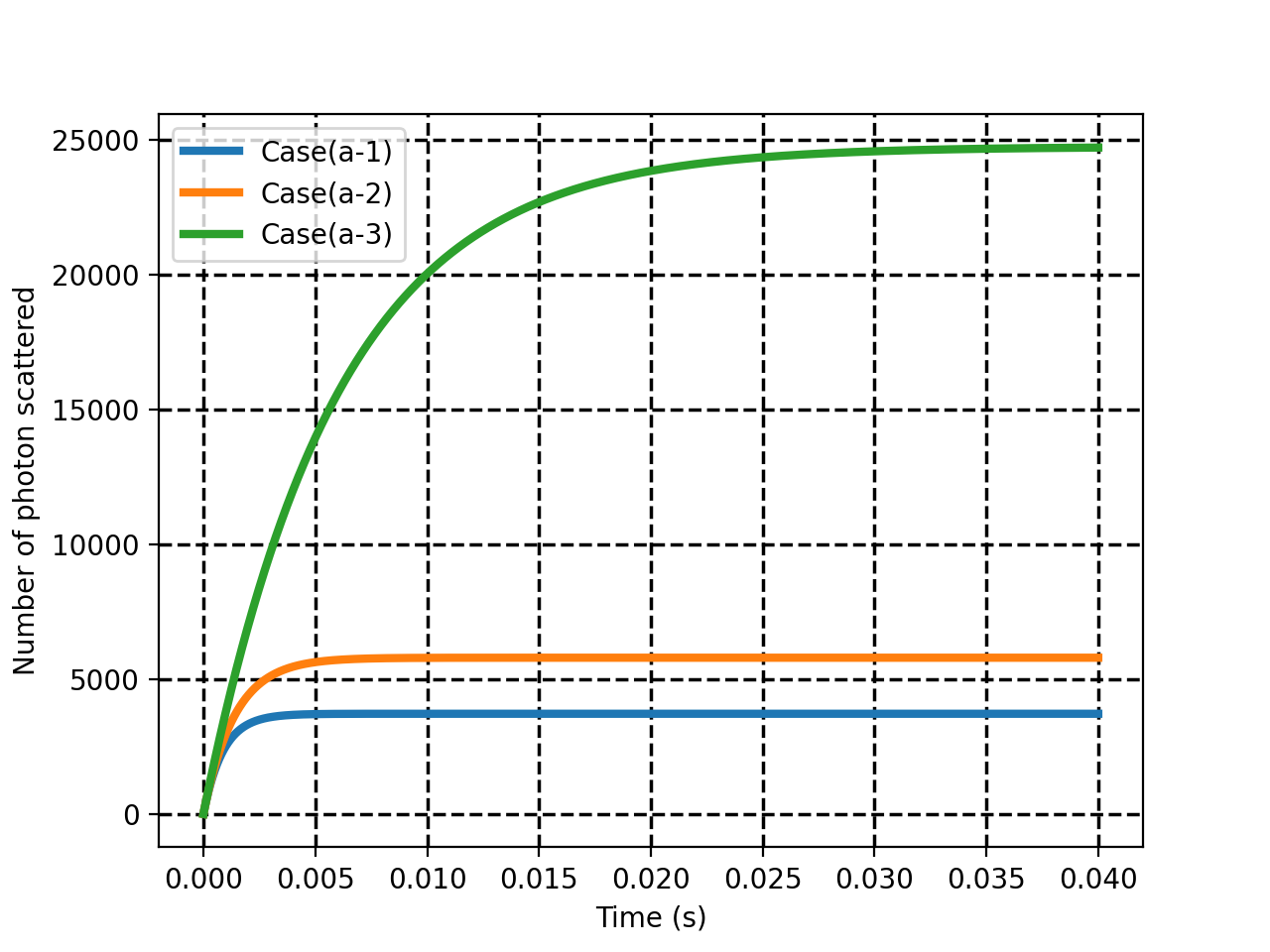}
\end{minipage}
}
\caption{\label{fig:epsart_simulation_1} The number of scattered photon for TlF and TlCl with different laser configurations. Case \textbf{(a-1)} has first three laser: $\lambda_{0'0''}$, $\lambda_{1'1''}$, and $\lambda_{0'2''}$. Case\textbf{(a-2)} includes Case \textbf{(a-1)}, plus the fourth laser $\lambda_{1'3''}$. Case \textbf{(a-3)} includes Case \textbf{(a-2)}, plus the fifth laser $\lambda_{2'4''}$.}
\label{fig}
\end{figure}

We use three different laser configurations in this simulation: Case \textbf{(a-1)} includes three laser: $\lambda_{0'0''}$, $\lambda_{1'1''}$, and $\lambda_{0'2''}$; Case \textbf{(a-2)} has an additional laser $\lambda_{1'3''}$. Case \textbf{(a-3)} contains all five lasers. The simulation results are plotted in Figure \ref{fig:epsart_simulation_1}. The population is initially in X$^1\Sigma_{0}^{+}$ (v''=0) state. 

The dynamics of this two molecules are similar. It is seen from the figure that TlF arrives at the corresponding limit faster than TlCl since its spontaneous radiation rate is almost twice as much as rate of TlCl. On the other hand, we note that TlCl  scatters more photons during the cooling process than TlF does. Specially, in the situation of employing five lasers, TlF absorbs about 7300 photons but TlCl absorbs 25000 photons in this model. 

A simple equation 
\begin{equation}
N_{tot}=\frac{1}{1-\sum_{i=0}^{i=4}R_{0'i''}}
\end{equation} 
could be used to qualitatively estimate the total photon absorption/emission cycles\cite{fu2017laser}: 
It's plain to see $N_{tot}$ is sensitive on the vibrational branching, particularly on the non-diagonal element in Franck-Condon factors such as (v'=0) $\to$ (v''=1,2,3...etc). Such sensitivity is also presented on the significant difference between five laser and four laser configuration. 

The less scattered photons means that we need to increase the number of laser for sufficiently cooling. The accurate calculation of non-diagonal value of FCFs would require a higher level of accuracy in the electron correlation calculations, such as by the inclusion of  triple (T) and higher excitation in CI and coupled-cluster wave-function, or to consider the non-adiabatic corrections. Due to constraints in computational resources and availability of computer implementations, the exploration of such factors is beyond the scope of the current work. Before delving into that, however, it would be interesting to have accurate experimental data on vibrational branching measurement, in order to gauge how much theory will have to improve to bridge the gap to experiment for TlCl.

\begin{figure}  [htb]
\centering
\subfigure [TlF]{
\label{fig:a}
\begin{minipage}[b]{0.47\textwidth}
\includegraphics [width=\linewidth] {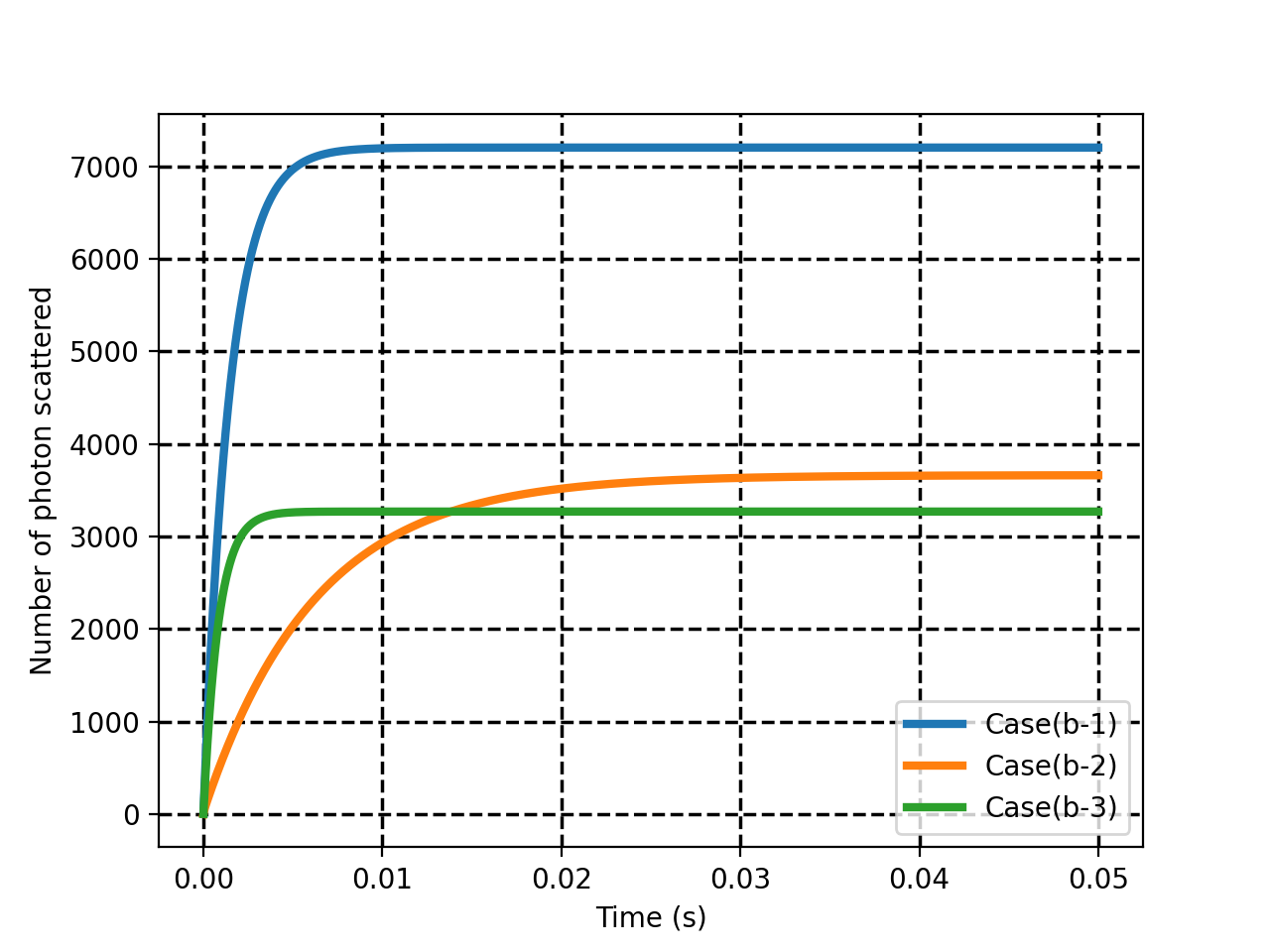}
\end{minipage}
}

\subfigure [TlCl]{
\label{fig:b}
\begin{minipage}[b]{0.47\textwidth}
\includegraphics [width=\linewidth] {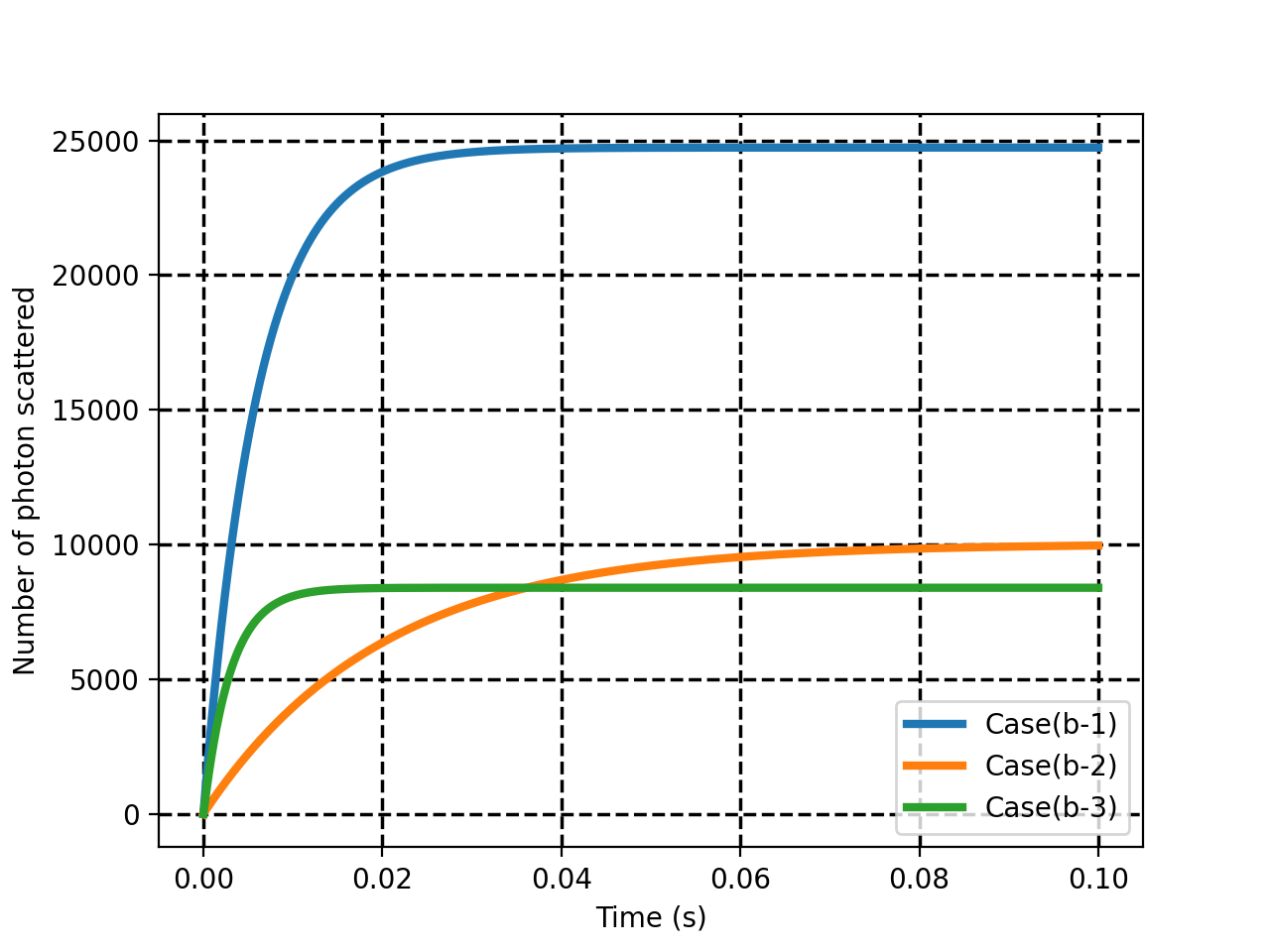}
\end{minipage}
}
\caption{\label{fig:epsart_diffe_sti} The cooling simulation at different levels of stimulated radiation. Case \textbf{(b-1)}: spectral energy density $\rho(\omega_{ij})$ is $10^{-12}$J/(m$^{3}\cdot$s$\cdot$Hz) ; Case \textbf{(b-2)}: $\rho(\omega_{ij})$ is $10^{-13}$J/(m$^{3}\cdot$s$\cdot$Hz) ; Case \textbf{(b-3)}: $\rho(\omega_{ij})$ is $10^{-12}$J/(m$^{3}\cdot$s$\cdot$Hz)  but the stimulated radiation coefficients $B_{ij}$ is set as 0.}
\label{fig}
\end{figure}

In addition, as reported by Norcia et al\cite{norcia2018narrow}, introducing the stimulated emission is a potential efficient method for laser cooling. Due to that, here we also investigate the effect of stimulated radiation by changing the spectral energy density $\rho(\omega_{ij})$ of the simulation in which employs five lasers. The $\rho(\omega_{ij})$ in Case \textbf{(b-1)} and Case \textbf{(b-2)} are $10^{-12}$ J/(m$^{3}\cdot$s$\cdot$Hz) and $10^{-13}$ J/(m$^{3}\cdot$s$\cdot$Hz) respectively. In Case \textbf{(b-3)}, we remove the stimulated radiation terms in the equation with keeping the same $\rho(\omega_{ij})$ as  Case \textbf{(b-1)}. The results are displayed on Figure \ref{fig:epsart_diffe_sti}. The large difference on both photons number and scattering rate between Case \textbf{(b-1)} and Case \textbf{(b-3)} shows that the influence of stimulated radiation is significant in the simulation.  

By comparing with Case \textbf{(b-1)} and Case \textbf{(b-2)}, we find that increasing higher spectral energy density results in absorbing more photons. For instance, TlCl could be scattered 25000 photons in 0.04 s under $\rho(\omega_{ij}) = 10^{-12}$ J/(m$^{3}\cdot$s$\cdot$Hz) , but only 10000 in 0.08s under $\rho(\omega_{ij}) = 10^{-13}$ J/(m$^{3}\cdot$s$\cdot$Hz) . In conclusion, enhancing the stimulated radiation is a good method for improving the cooling property including both total number of scattered photon and scattering rate.

\section{Conclusion}

In this work we have investigated the permanent dipole moments (PDMs) for ground and low-lying excited states of TlF and TlCl molecules, as well as transition dipole moments (TDMs) between these electronic states, via four-component Multi-reference Configuration Interaction, equation of motion coupled-cluster and polarization propagator calculations. Our main goal is to obtain, from the TDMs, the excited state lifetimes that allow us to to predict whether the TlCl species is a suitable candidate for laser cooling experiments. 

After cross-validating the four-component MRCI results with the other two approaches, we have employed it to obtain a PDM of -2.47 D and lifetime 91 ns of a$^{3}\Pi_{1}$ state of TlF, which are close to the experimental results of -2.28(7) D and 99(9) ns, respectively. 

For TlCl, we obtained from our four-component MRCI calculations a lifetime of 175 ns for the a$^{3}\Pi_{0}^{+}$ state. This value, which is much shorter than a recent theoretical estimation of 6.04 $\mu s$, obtained from spin-orbit CI calculations. Our results point to the strong underestimation of the TDMs as the main factor behind such a discrepancy, as the SOCI ground and excited state energies and PDMs for TlCl closely match the four-component values.

With the new lifetime, we have performed a new a population simulation by solving the rate equation, and we find that TlCl shows cooling dynamics similar to that of TlF. Our simulations also shows that the vibrational branching of weak transitions determined by non-diagonal element of Franck-Condon factors is potentially very important for cooling efficiency. Because of this, we consider that highly accurate experiment on Franck-Condon factor of  a$^{3}\Pi_{0}^{+}$-X$^{1}\Sigma^{+}$ transition of TlCl could provide useful information for refining theoretical models.

Finally, we analyzed the influence of stimulated radiation on cooling process. We show the stimulated radiation is important and increasing the spectral energy density is a possible way to improve the cooling efficiency.

\begin{acknowledgments}
We acknowledge funding from projects CompRIXS (ANR-19-CE29-0019, DFG JA 2329/6-1), Labex CaPPA (ANR-11-LABX-0005-01) and the I-SITE ULNE project OVERSEE and MESONM International Associated Laboratory (LAI) (ANR-16-IDEX-0004), and support from the French national supercomputing facilities (grant DARI A0090801859).
\end{acknowledgments}

\bibliographystyle{aipnum4-1}
\bibliography{aipsamp.bib}

\providecommand{\noopsort}[1]{}\providecommand{\singleletter}[1]{#1}%
\begin{thebibliography}{42}%
\makeatletter
\providecommand \@ifxundefined [1]{%
 \@ifx{#1\undefined}
}%
\providecommand \@ifnum [1]{%
 \ifnum #1\expandafter \@firstoftwo
 \else \expandafter \@secondoftwo
 \fi
}%
\providecommand \@ifx [1]{%
 \ifx #1\expandafter \@firstoftwo
 \else \expandafter \@secondoftwo
 \fi
}%
\providecommand \natexlab [1]{#1}%
\providecommand \enquote  [1]{``#1''}%
\providecommand \bibnamefont  [1]{#1}%
\providecommand \bibfnamefont [1]{#1}%
\providecommand \citenamefont [1]{#1}%
\providecommand \href@noop [0]{\@secondoftwo}%
\providecommand \href [0]{\begingroup \@sanitize@url \@href}%
\providecommand \@href[1]{\@@startlink{#1}\@@href}%
\providecommand \@@href[1]{\endgroup#1\@@endlink}%
\providecommand \@sanitize@url [0]{\catcode `\\12\catcode `\$12\catcode
  `\&12\catcode `\#12\catcode `\^12\catcode `\_12\catcode `\%12\relax}%
\providecommand \@@startlink[1]{}%
\providecommand \@@endlink[0]{}%
\providecommand \url  [0]{\begingroup\@sanitize@url \@url }%
\providecommand \@url [1]{\endgroup\@href {#1}{\urlprefix }}%
\providecommand \urlprefix  [0]{URL }%
\providecommand \Eprint [0]{\href }%
\providecommand \doibase [0]{http://dx.doi.org/}%
\providecommand \selectlanguage [0]{\@gobble}%
\providecommand \bibinfo  [0]{\@secondoftwo}%
\providecommand \bibfield  [0]{\@secondoftwo}%
\providecommand \translation [1]{[#1]}%
\providecommand \BibitemOpen [0]{}%
\providecommand \bibitemStop [0]{}%
\providecommand \bibitemNoStop [0]{.\EOS\space}%
\providecommand \EOS [0]{\spacefactor3000\relax}%
\providecommand \BibitemShut  [1]{\csname bibitem#1\endcsname}%
\let\auto@bib@innerbib\@empty
\bibitem [{\citenamefont {Safronova}\ \emph {et~al.}(2018)\citenamefont
  {Safronova}, \citenamefont {Budker}, \citenamefont {DeMille}, \citenamefont
  {Kimball}, \citenamefont {Derevianko},\ and\ \citenamefont
  {Clark}}]{safronova2018search}%
  \BibitemOpen
  \bibfield  {author} {\bibinfo {author} {\bibfnamefont {M.~S.}\ \bibnamefont
  {Safronova}}, \bibinfo {author} {\bibfnamefont {D.}~\bibnamefont {Budker}},
  \bibinfo {author} {\bibfnamefont {D.}~\bibnamefont {DeMille}}, \bibinfo
  {author} {\bibfnamefont {D.~F.~J.}\ \bibnamefont {Kimball}}, \bibinfo
  {author} {\bibfnamefont {A.}~\bibnamefont {Derevianko}}, \ and\ \bibinfo
  {author} {\bibfnamefont {C.~W.}\ \bibnamefont {Clark}},\ }\href@noop {}
  {\bibfield  {journal} {\bibinfo  {journal} {Rev. Mod. Phys}\ }\textbf
  {\bibinfo {volume} {90}},\ \bibinfo {pages} {025008} (\bibinfo {year}
  {2018})}\BibitemShut {NoStop}%
\bibitem [{\citenamefont {Rosa}(2004)}]{di2004laser}%
  \BibitemOpen
  \bibfield  {author} {\bibinfo {author} {\bibfnamefont {M.~D.~D.}\
  \bibnamefont {Rosa}},\ }\href@noop {} {\bibfield  {journal} {\bibinfo
  {journal} {Eur. Phys. J. D}\ }\textbf {\bibinfo {volume} {31}},\ \bibinfo
  {pages} {395} (\bibinfo {year} {2004})}\BibitemShut {NoStop}%
\bibitem [{\citenamefont {Shuman}, \citenamefont {Barry},\ and\ \citenamefont
  {DeMille}(2010)}]{shuman2010laser}%
  \BibitemOpen
  \bibfield  {author} {\bibinfo {author} {\bibfnamefont {E.~S.}\ \bibnamefont
  {Shuman}}, \bibinfo {author} {\bibfnamefont {J.~F.}\ \bibnamefont {Barry}}, \
  and\ \bibinfo {author} {\bibfnamefont {D.}~\bibnamefont {DeMille}},\
  }\href@noop {} {\bibfield  {journal} {\bibinfo  {journal} {Nature}\ }\textbf
  {\bibinfo {volume} {467}},\ \bibinfo {pages} {820} (\bibinfo {year}
  {2010})}\BibitemShut {NoStop}%
\bibitem [{\citenamefont {Truppe}\ \emph {et~al.}(2017)\citenamefont {Truppe},
  \citenamefont {Williams}, \citenamefont {Hambach}, \citenamefont {Caldwell},
  \citenamefont {Fitch}, \citenamefont {Hinds}, \citenamefont {Sauer},\ and\
  \citenamefont {Tarbutt}}]{truppe2017molecules}%
  \BibitemOpen
  \bibfield  {author} {\bibinfo {author} {\bibfnamefont {S.}~\bibnamefont
  {Truppe}}, \bibinfo {author} {\bibfnamefont {H.~J.}\ \bibnamefont
  {Williams}}, \bibinfo {author} {\bibfnamefont {M.}~\bibnamefont {Hambach}},
  \bibinfo {author} {\bibfnamefont {L.}~\bibnamefont {Caldwell}}, \bibinfo
  {author} {\bibfnamefont {N.~J.}\ \bibnamefont {Fitch}}, \bibinfo {author}
  {\bibfnamefont {E.~A.}\ \bibnamefont {Hinds}}, \bibinfo {author}
  {\bibfnamefont {B.~E.}\ \bibnamefont {Sauer}}, \ and\ \bibinfo {author}
  {\bibfnamefont {M.~R.}\ \bibnamefont {Tarbutt}},\ }\href@noop {} {\bibfield
  {journal} {\bibinfo  {journal} {Nat. Phys}\ }\textbf {\bibinfo {volume}
  {13}},\ \bibinfo {pages} {1173} (\bibinfo {year} {2017})}\BibitemShut
  {NoStop}%
\bibitem [{\citenamefont {Hummon}\ \emph {et~al.}(2013)\citenamefont {Hummon},
  \citenamefont {Yeo}, \citenamefont {Stuhl}, \citenamefont {Collopy},
  \citenamefont {Xia},\ and\ \citenamefont {Ye}}]{hummon20132d}%
  \BibitemOpen
  \bibfield  {author} {\bibinfo {author} {\bibfnamefont {M.~T.}\ \bibnamefont
  {Hummon}}, \bibinfo {author} {\bibfnamefont {M.}~\bibnamefont {Yeo}},
  \bibinfo {author} {\bibfnamefont {B.~K.}\ \bibnamefont {Stuhl}}, \bibinfo
  {author} {\bibfnamefont {A.~L.}\ \bibnamefont {Collopy}}, \bibinfo {author}
  {\bibfnamefont {Y.}~\bibnamefont {Xia}}, \ and\ \bibinfo {author}
  {\bibfnamefont {J.}~\bibnamefont {Ye}},\ }\href@noop {} {\bibfield  {journal}
  {\bibinfo  {journal} {Phys. Rev. Lett}\ }\textbf {\bibinfo {volume} {110}},\
  \bibinfo {pages} {143001} (\bibinfo {year} {2013})}\BibitemShut {NoStop}%
\bibitem [{\citenamefont {Lim}\ \emph {et~al.}(2018)\citenamefont {Lim},
  \citenamefont {Almond}, \citenamefont {Trigatzis}, \citenamefont {Devlin},
  \citenamefont {Fitch}, \citenamefont {Sauer}, \citenamefont {Tarbutt},\ and\
  \citenamefont {Hinds}}]{lim2018laser}%
  \BibitemOpen
  \bibfield  {author} {\bibinfo {author} {\bibfnamefont {J.}~\bibnamefont
  {Lim}}, \bibinfo {author} {\bibfnamefont {J.~R.}\ \bibnamefont {Almond}},
  \bibinfo {author} {\bibfnamefont {M.~A.}\ \bibnamefont {Trigatzis}}, \bibinfo
  {author} {\bibfnamefont {J.~A.}\ \bibnamefont {Devlin}}, \bibinfo {author}
  {\bibfnamefont {N.~J.}\ \bibnamefont {Fitch}}, \bibinfo {author}
  {\bibfnamefont {B.~E.}\ \bibnamefont {Sauer}}, \bibinfo {author}
  {\bibfnamefont {M.~R.}\ \bibnamefont {Tarbutt}}, \ and\ \bibinfo {author}
  {\bibfnamefont {E.~A.}\ \bibnamefont {Hinds}},\ }\href@noop {} {\bibfield
  {journal} {\bibinfo  {journal} {Phys. Rev. Lett}\ }\textbf {\bibinfo {volume}
  {120}},\ \bibinfo {pages} {123201} (\bibinfo {year} {2018})}\BibitemShut
  {NoStop}%
\bibitem [{\citenamefont {Hinds}\ and\ \citenamefont
  {Sandars}(1980)}]{hinds1980experiment}%
  \BibitemOpen
  \bibfield  {author} {\bibinfo {author} {\bibfnamefont {E.~A.}\ \bibnamefont
  {Hinds}}\ and\ \bibinfo {author} {\bibfnamefont {P.~G.~H.}\ \bibnamefont
  {Sandars}},\ }\href@noop {} {\bibfield  {journal} {\bibinfo  {journal} {Phys.
  Rev. A}\ }\textbf {\bibinfo {volume} {21}},\ \bibinfo {pages} {480} (\bibinfo
  {year} {1980})}\BibitemShut {NoStop}%
\bibitem [{\citenamefont {Sandars}(1967)}]{sandars_measurability_1967}%
  \BibitemOpen
  \bibfield  {author} {\bibinfo {author} {\bibfnamefont {P.~G.~H.}\
  \bibnamefont {Sandars}},\ }\href {\doibase 10.1103/PhysRevLett.19.1396}
  {\bibfield  {journal} {\bibinfo  {journal} {Phys. Rev. Lett.}\ }\textbf
  {\bibinfo {volume} {19}},\ \bibinfo {pages} {1396} (\bibinfo {year}
  {1967})}\BibitemShut {NoStop}%
\bibitem [{\citenamefont {Hunter}\ \emph {et~al.}(2012)\citenamefont {Hunter},
  \citenamefont {Peck}, \citenamefont {Greenspon}, \citenamefont {Alam},\ and\
  \citenamefont {DeMille}}]{hunter2012prospects}%
  \BibitemOpen
  \bibfield  {author} {\bibinfo {author} {\bibfnamefont {L.~R.}\ \bibnamefont
  {Hunter}}, \bibinfo {author} {\bibfnamefont {S.~K.}\ \bibnamefont {Peck}},
  \bibinfo {author} {\bibfnamefont {A.~S.}\ \bibnamefont {Greenspon}}, \bibinfo
  {author} {\bibfnamefont {S.~S.}\ \bibnamefont {Alam}}, \ and\ \bibinfo
  {author} {\bibfnamefont {D.}~\bibnamefont {DeMille}},\ }\href@noop {}
  {\bibfield  {journal} {\bibinfo  {journal} {Phys. Rev. A}\ }\textbf {\bibinfo
  {volume} {85}},\ \bibinfo {pages} {012511} (\bibinfo {year}
  {2012})}\BibitemShut {NoStop}%
\bibitem [{\citenamefont {Clayburn}\ \emph {et~al.}(2020)\citenamefont
  {Clayburn}, \citenamefont {Wright}, \citenamefont {Norrgard}, \citenamefont
  {DeMille},\ and\ \citenamefont {Hunter}}]{clayburn2020measurement}%
  \BibitemOpen
  \bibfield  {author} {\bibinfo {author} {\bibfnamefont {N.~B.}\ \bibnamefont
  {Clayburn}}, \bibinfo {author} {\bibfnamefont {T.~H.}\ \bibnamefont
  {Wright}}, \bibinfo {author} {\bibfnamefont {E.~B.}\ \bibnamefont
  {Norrgard}}, \bibinfo {author} {\bibfnamefont {D.}~\bibnamefont {DeMille}}, \
  and\ \bibinfo {author} {\bibfnamefont {L.~R.}\ \bibnamefont {Hunter}},\
  }\href@noop {} {\bibfield  {journal} {\bibinfo  {journal} {Phys. Rev. A}\
  }\textbf {\bibinfo {volume} {102}},\ \bibinfo {pages} {052802} (\bibinfo
  {year} {2020})}\BibitemShut {NoStop}%
\bibitem [{\citenamefont {Meijer}\ and\ \citenamefont
  {Sartakov}(2020)}]{meijer2020lambda}%
  \BibitemOpen
  \bibfield  {author} {\bibinfo {author} {\bibfnamefont {G.}~\bibnamefont
  {Meijer}}\ and\ \bibinfo {author} {\bibfnamefont {B.~G.}\ \bibnamefont
  {Sartakov}},\ }\href@noop {} {\bibfield  {journal} {\bibinfo  {journal}
  {Phys. Rev. A}\ }\textbf {\bibinfo {volume} {101}},\ \bibinfo {pages}
  {042506} (\bibinfo {year} {2020})}\BibitemShut {NoStop}%
\bibitem [{\citenamefont {Zou}\ and\ \citenamefont
  {Liu}(2009)}]{Zou2009Comprehensive}%
  \BibitemOpen
  \bibfield  {author} {\bibinfo {author} {\bibfnamefont {W.}~\bibnamefont
  {Zou}}\ and\ \bibinfo {author} {\bibfnamefont {W.}~\bibnamefont {Liu}},\
  }\href@noop {} {\bibfield  {journal} {\bibinfo  {journal} {J. Com. Chem}\
  }\textbf {\bibinfo {volume} {30}},\ \bibinfo {pages} {524–539} (\bibinfo
  {year} {2009})}\BibitemShut {NoStop}%
\bibitem [{\citenamefont {Yuan}\ \emph {et~al.}(2018)\citenamefont {Yuan},
  \citenamefont {Yin}, \citenamefont {Shen}, \citenamefont {Liu}, \citenamefont
  {Lian}, \citenamefont {Xu},\ and\ \citenamefont {Yan}}]{yuan2018laser}%
  \BibitemOpen
  \bibfield  {author} {\bibinfo {author} {\bibfnamefont {X.}~\bibnamefont
  {Yuan}}, \bibinfo {author} {\bibfnamefont {S.}~\bibnamefont {Yin}}, \bibinfo
  {author} {\bibfnamefont {Y.}~\bibnamefont {Shen}}, \bibinfo {author}
  {\bibfnamefont {Y.}~\bibnamefont {Liu}}, \bibinfo {author} {\bibfnamefont
  {Y.}~\bibnamefont {Lian}}, \bibinfo {author} {\bibfnamefont {H.}~\bibnamefont
  {Xu}}, \ and\ \bibinfo {author} {\bibfnamefont {B.}~\bibnamefont {Yan}},\
  }\href@noop {} {\bibfield  {journal} {\bibinfo  {journal} {J. Chem. Phys}\
  }\textbf {\bibinfo {volume} {149}},\ \bibinfo {pages} {094306} (\bibinfo
  {year} {2018})}\BibitemShut {NoStop}%
\bibitem [{\citenamefont {Li}\ \emph {et~al.}(1994)\citenamefont {Li},
  \citenamefont {Liebermann}, \citenamefont {Hirsch},\ and\ \citenamefont
  {Buenker}}]{li1994relativistic}%
  \BibitemOpen
  \bibfield  {author} {\bibinfo {author} {\bibfnamefont {Y.}~\bibnamefont
  {Li}}, \bibinfo {author} {\bibfnamefont {H.~P.}\ \bibnamefont {Liebermann}},
  \bibinfo {author} {\bibfnamefont {G.}~\bibnamefont {Hirsch}}, \ and\ \bibinfo
  {author} {\bibfnamefont {R.~J.}\ \bibnamefont {Buenker}},\ }\href@noop {}
  {\bibfield  {journal} {\bibinfo  {journal} {J. Mol. Spectrosc}\ }\textbf
  {\bibinfo {volume} {165}},\ \bibinfo {pages} {219} (\bibinfo {year}
  {1994})}\BibitemShut {NoStop}%
\bibitem [{\citenamefont {Vallet}\ \emph {et~al.}(2000)\citenamefont {Vallet},
  \citenamefont {Maron}, \citenamefont {Teichteil},\ and\ \citenamefont
  {Flament}}]{Vallet2000}%
  \BibitemOpen
  \bibfield  {author} {\bibinfo {author} {\bibfnamefont {V.}~\bibnamefont
  {Vallet}}, \bibinfo {author} {\bibfnamefont {L.}~\bibnamefont {Maron}},
  \bibinfo {author} {\bibfnamefont {C.}~\bibnamefont {Teichteil}}, \ and\
  \bibinfo {author} {\bibfnamefont {J.-P.}\ \bibnamefont {Flament}},\ }\href
  {\doibase 10.1063/1.481929} {\bibfield  {journal} {\bibinfo  {journal} {J.
  Chem. Phys.}\ }\textbf {\bibinfo {volume} {113}},\ \bibinfo {pages} {1391}
  (\bibinfo {year} {2000})}\BibitemShut {NoStop}%
\bibitem [{\citenamefont {Weigand}\ \emph {et~al.}(2009)\citenamefont
  {Weigand}, \citenamefont {Cao}, \citenamefont {Vallet}, \citenamefont
  {Flament},\ and\ \citenamefont {Dolg}}]{Weigand:2009dq}%
  \BibitemOpen
  \bibfield  {author} {\bibinfo {author} {\bibfnamefont {A.}~\bibnamefont
  {Weigand}}, \bibinfo {author} {\bibfnamefont {X.}~\bibnamefont {Cao}},
  \bibinfo {author} {\bibfnamefont {V.}~\bibnamefont {Vallet}}, \bibinfo
  {author} {\bibfnamefont {J.-P.}\ \bibnamefont {Flament}}, \ and\ \bibinfo
  {author} {\bibfnamefont {M.}~\bibnamefont {Dolg}},\ }\href {\doibase
  10.1021/jp902693b} {\bibfield  {journal} {\bibinfo  {journal} {J. Phys. Chem.
  A}\ }\textbf {\bibinfo {volume} {113}},\ \bibinfo {pages} {11509} (\bibinfo
  {year} {2009})}\BibitemShut {NoStop}%
\bibitem [{\citenamefont {Danilo}\ \emph {et~al.}(2010)\citenamefont {Danilo},
  \citenamefont {Vallet}, \citenamefont {Flament},\ and\ \citenamefont
  {Wahlgren}}]{Danilo2010}%
  \BibitemOpen
  \bibfield  {author} {\bibinfo {author} {\bibfnamefont {C.}~\bibnamefont
  {Danilo}}, \bibinfo {author} {\bibfnamefont {V.}~\bibnamefont {Vallet}},
  \bibinfo {author} {\bibfnamefont {J.-P.}\ \bibnamefont {Flament}}, \ and\
  \bibinfo {author} {\bibfnamefont {U.}~\bibnamefont {Wahlgren}},\ }\href
  {http://dx.doi.org/10.1039/B914222C} {\bibfield  {journal} {\bibinfo
  {journal} {Phys. Chem. Chem. Phys.}\ }\textbf {\bibinfo {volume} {12}},\
  \bibinfo {pages} {1116} (\bibinfo {year} {2010})}\BibitemShut {NoStop}%
\bibitem [{\citenamefont {Gomes}\ \emph {et~al.}(2014)\citenamefont {Gomes},
  \citenamefont {R{\'e}al}, \citenamefont {Galland}, \citenamefont {Angeli},
  \citenamefont {Cimiraglia},\ and\ \citenamefont
  {Vallet}}]{PereiraGomes:2014iy}%
  \BibitemOpen
  \bibfield  {author} {\bibinfo {author} {\bibfnamefont {A.~S.~P.}\
  \bibnamefont {Gomes}}, \bibinfo {author} {\bibfnamefont {F.}~\bibnamefont
  {R{\'e}al}}, \bibinfo {author} {\bibfnamefont {N.}~\bibnamefont {Galland}},
  \bibinfo {author} {\bibfnamefont {C.}~\bibnamefont {Angeli}}, \bibinfo
  {author} {\bibfnamefont {R.}~\bibnamefont {Cimiraglia}}, \ and\ \bibinfo
  {author} {\bibfnamefont {V.}~\bibnamefont {Vallet}},\ }\href {\doibase
  10.1039/C3CP55294B} {\bibfield  {journal} {\bibinfo  {journal} {Phys. Chem.
  Chem. Phys.}\ }\textbf {\bibinfo {volume} {16}},\ \bibinfo {pages} {9238}
  (\bibinfo {year} {2014})}\BibitemShut {NoStop}%
\bibitem [{\citenamefont {Kervazo}\ \emph {et~al.}(2019)\citenamefont
  {Kervazo}, \citenamefont {Réal}, \citenamefont {Virot}, \citenamefont
  {Severo Pereira~Gomes},\ and\ \citenamefont {Vallet}}]{Kervazo2019}%
  \BibitemOpen
  \bibfield  {author} {\bibinfo {author} {\bibfnamefont {S.}~\bibnamefont
  {Kervazo}}, \bibinfo {author} {\bibfnamefont {F.}~\bibnamefont {Réal}},
  \bibinfo {author} {\bibfnamefont {F.}~\bibnamefont {Virot}}, \bibinfo
  {author} {\bibfnamefont {A.}~\bibnamefont {Severo Pereira~Gomes}}, \ and\
  \bibinfo {author} {\bibfnamefont {V.}~\bibnamefont {Vallet}},\ }\href
  {\doibase 10.1021/acs.inorgchem.9b02096} {\bibfield  {journal} {\bibinfo
  {journal} {Inorg. Chem.}\ }\textbf {\bibinfo {volume} {58}},\ \bibinfo
  {pages} {14507} (\bibinfo {year} {2019})}\BibitemShut {NoStop}%
\bibitem [{\citenamefont {Bartlett}(2012)}]{bartlett_coupledcluster_2012}%
  \BibitemOpen
  \bibfield  {author} {\bibinfo {author} {\bibfnamefont {R.~J.}\ \bibnamefont
  {Bartlett}},\ }\href {\doibase 10.1002/wcms.76} {\bibfield  {journal}
  {\bibinfo  {journal} {WIREs Comput Mol Sci}\ }\textbf {\bibinfo {volume}
  {2}},\ \bibinfo {pages} {126} (\bibinfo {year} {2012})}\BibitemShut {NoStop}%
\bibitem [{\citenamefont {Dreuw}\ and\ \citenamefont
  {Wormit}(2015)}]{dreuw_algebraic_2015}%
  \BibitemOpen
  \bibfield  {author} {\bibinfo {author} {\bibfnamefont {A.}~\bibnamefont
  {Dreuw}}\ and\ \bibinfo {author} {\bibfnamefont {M.}~\bibnamefont {Wormit}},\
  }\href {\doibase 10.1002/wcms.1206} {\bibfield  {journal} {\bibinfo
  {journal} {WIREs Comput Mol Sci}\ }\textbf {\bibinfo {volume} {5}},\ \bibinfo
  {pages} {82} (\bibinfo {year} {2015})}\BibitemShut {NoStop}%
\bibitem [{DIR()}]{DIRAC19}%
  \BibitemOpen
  \href@noop {} {}\bibinfo {note} {{DIRAC}, a relativistic ab initio electronic
  structure program, Release {DIRAC19} (2019), written by A.~S.~P.~Gomes,
  T.~Saue, L.~Visscher, H.~J.~{\relax Aa}.~Jensen, and R.~Bast, with
  contributions from I.~A.~Aucar, V.~Bakken, K.~G.~Dyall, S.~Dubillard,
  U.~Ekstr{\"o}m, E.~Eliav, T.~Enevoldsen, E.~Fa{\ss}hauer, T.~Fleig,
  O.~Fossgaard, L.~Halbert, E.~D.~Hedeg{\aa}rd, B.~Heimlich--Paris,
  T.~Helgaker, J.~Henriksson, M.~Ilia{\v{s}}, Ch.~R.~Jacob, S.~Knecht,
  S.~Komorovsk{\'y}, O.~Kullie, J.~K.~L{\ae}rdahl, C.~V.~Larsen, Y.~S.~Lee,
  H.~S.~Nataraj, M.~K.~Nayak, P.~Norman, G.~Olejniczak, J.~Olsen,
  J.~M.~H.~Olsen, Y.~C.~Park, J.~K.~Pedersen, M.~Pernpointner, R.~di~Remigio,
  K.~Ruud, P.~Sa{\l}ek, B.~Schimmelpfennig, B.~Senjean, A.~Shee, J.~Sikkema,
  A.~J.~Thorvaldsen, J.~Thyssen, J.~van~Stralen, M.~L.~Vidal, S.~Villaume,
  O.~Visser, T.~Winther, and S.~Yamamoto (available at
  \url{http://dx.doi.org/10.5281/zenodo.3572669}, see also
  \url{http://www.diracprogram.org})}\BibitemShut {NoStop}%
\bibitem [{\citenamefont {Saue}\ \emph {et~al.}(2020)\citenamefont {Saue},
  \citenamefont {Bast}, \citenamefont {Gomes}, \citenamefont {Jensen},
  \citenamefont {Visscher}, \citenamefont {Aucar}, \citenamefont {Di},
  \citenamefont {Dyall}, \citenamefont {Eliav}, \citenamefont {Fasshauer} \emph
  {et~al.}}]{saue2020dirac}%
  \BibitemOpen
  \bibfield  {author} {\bibinfo {author} {\bibfnamefont {T.}~\bibnamefont
  {Saue}}, \bibinfo {author} {\bibfnamefont {R.}~\bibnamefont {Bast}}, \bibinfo
  {author} {\bibfnamefont {A.~S.~P.}\ \bibnamefont {Gomes}}, \bibinfo {author}
  {\bibfnamefont {H.~J.~A.}\ \bibnamefont {Jensen}}, \bibinfo {author}
  {\bibfnamefont {L.}~\bibnamefont {Visscher}}, \bibinfo {author}
  {\bibfnamefont {L.~A.}\ \bibnamefont {Aucar}}, \bibinfo {author}
  {\bibfnamefont {R.~R.}\ \bibnamefont {Di}}, \bibinfo {author} {\bibfnamefont
  {K.~G.}\ \bibnamefont {Dyall}}, \bibinfo {author} {\bibfnamefont
  {E.}~\bibnamefont {Eliav}}, \bibinfo {author} {\bibfnamefont
  {E.}~\bibnamefont {Fasshauer}},  \emph {et~al.},\ }\href@noop {} {\bibfield
  {journal} {\bibinfo  {journal} {J. Chem. Phys}\ }\textbf {\bibinfo {volume}
  {152}},\ \bibinfo {pages} {204104} (\bibinfo {year} {2020})}\BibitemShut
  {NoStop}%
\bibitem [{\citenamefont {Visscher}(1997)}]{visscher1997approximate}%
  \BibitemOpen
  \bibfield  {author} {\bibinfo {author} {\bibfnamefont {L.}~\bibnamefont
  {Visscher}},\ }\href@noop {} {\bibfield  {journal} {\bibinfo  {journal}
  {Theor. Chem. Acc}\ }\textbf {\bibinfo {volume} {98}},\ \bibinfo {pages} {68}
  (\bibinfo {year} {1997})}\BibitemShut {NoStop}%
\bibitem [{\citenamefont {Dyall}(2006)}]{dyall2006relativistic}%
  \BibitemOpen
  \bibfield  {author} {\bibinfo {author} {\bibfnamefont {K.~G.}\ \bibnamefont
  {Dyall}},\ }\href@noop {} {\bibfield  {journal} {\bibinfo  {journal} {Theor.
  Chem. Acc}\ }\textbf {\bibinfo {volume} {115}},\ \bibinfo {pages} {441}
  (\bibinfo {year} {2006})}\BibitemShut {NoStop}%
\bibitem [{\citenamefont {Kendall}, \citenamefont {Jr},\ and\ \citenamefont
  {Harrison}(1992)}]{kendall1992electron}%
  \BibitemOpen
  \bibfield  {author} {\bibinfo {author} {\bibfnamefont {R.~A.}\ \bibnamefont
  {Kendall}}, \bibinfo {author} {\bibfnamefont {T.~H.~D.}\ \bibnamefont {Jr}},
  \ and\ \bibinfo {author} {\bibfnamefont {R.~J.}\ \bibnamefont {Harrison}},\
  }\href@noop {} {\bibfield  {journal} {\bibinfo  {journal} {J. Chem. Phys}\
  }\textbf {\bibinfo {volume} {96}},\ \bibinfo {pages} {6796} (\bibinfo {year}
  {1992})}\BibitemShut {NoStop}%
\bibitem [{\citenamefont {Woon}\ and\ \citenamefont
  {Jr}(1993)}]{woon1993gaussian}%
  \BibitemOpen
  \bibfield  {author} {\bibinfo {author} {\bibfnamefont {D.~E.}\ \bibnamefont
  {Woon}}\ and\ \bibinfo {author} {\bibfnamefont {T.~H.~D.}\ \bibnamefont
  {Jr}},\ }\href@noop {} {\bibfield  {journal} {\bibinfo  {journal} {J. Chem.
  Phys}\ }\textbf {\bibinfo {volume} {98}},\ \bibinfo {pages} {1358} (\bibinfo
  {year} {1993})}\BibitemShut {NoStop}%
\bibitem [{\citenamefont {Deng}, \citenamefont {Lian},\ and\ \citenamefont
  {Zou}(2017)}]{deng2017permanent}%
  \BibitemOpen
  \bibfield  {author} {\bibinfo {author} {\bibfnamefont {D.}~\bibnamefont
  {Deng}}, \bibinfo {author} {\bibfnamefont {Y.}~\bibnamefont {Lian}}, \ and\
  \bibinfo {author} {\bibfnamefont {W.}~\bibnamefont {Zou}},\ }\href@noop {}
  {\bibfield  {journal} {\bibinfo  {journal} {Chem. Phys. Lett}\ }\textbf
  {\bibinfo {volume} {688}},\ \bibinfo {pages} {33} (\bibinfo {year}
  {2017})}\BibitemShut {NoStop}%
\bibitem [{\citenamefont {Aquilante}\ \emph {et~al.}(2020)\citenamefont
  {Aquilante}, \citenamefont {Autschbach}, \citenamefont {Baiardi},
  \citenamefont {Battaglia}, \citenamefont {Borin}, \citenamefont {Chibotaru},
  \citenamefont {Conti}, \citenamefont {Vico}, \citenamefont {Delcey},
  \citenamefont {Galv{\'a}n} \emph {et~al.}}]{aquilante2020modern}%
  \BibitemOpen
  \bibfield  {author} {\bibinfo {author} {\bibfnamefont {F.}~\bibnamefont
  {Aquilante}}, \bibinfo {author} {\bibfnamefont {J.}~\bibnamefont
  {Autschbach}}, \bibinfo {author} {\bibfnamefont {A.}~\bibnamefont {Baiardi}},
  \bibinfo {author} {\bibfnamefont {S.}~\bibnamefont {Battaglia}}, \bibinfo
  {author} {\bibfnamefont {V.~A.}\ \bibnamefont {Borin}}, \bibinfo {author}
  {\bibfnamefont {L.~F.}\ \bibnamefont {Chibotaru}}, \bibinfo {author}
  {\bibfnamefont {I.}~\bibnamefont {Conti}}, \bibinfo {author} {\bibfnamefont
  {L.~D.}\ \bibnamefont {Vico}}, \bibinfo {author} {\bibfnamefont
  {M.}~\bibnamefont {Delcey}}, \bibinfo {author} {\bibfnamefont {I.~F.}\
  \bibnamefont {Galv{\'a}n}},  \emph {et~al.},\ }\href@noop {} {\bibfield
  {journal} {\bibinfo  {journal} {J. Chem. Phys}\ }\textbf {\bibinfo {volume}
  {152}},\ \bibinfo {pages} {214117} (\bibinfo {year} {2020})}\BibitemShut
  {NoStop}%
\bibitem [{\citenamefont {Knecht}, \citenamefont {Jensen},\ and\ \citenamefont
  {Fleig}(2010)}]{knecht2010large}%
  \BibitemOpen
  \bibfield  {author} {\bibinfo {author} {\bibfnamefont {S.}~\bibnamefont
  {Knecht}}, \bibinfo {author} {\bibfnamefont {H.~J.~A.}\ \bibnamefont
  {Jensen}}, \ and\ \bibinfo {author} {\bibfnamefont {T.}~\bibnamefont
  {Fleig}},\ }\href@noop {} {\bibfield  {journal} {\bibinfo  {journal} {J.
  Chem. Phys}\ }\textbf {\bibinfo {volume} {132}},\ \bibinfo {pages} {014108}
  (\bibinfo {year} {2010})}\BibitemShut {NoStop}%
\bibitem [{\citenamefont {Fleig}, \citenamefont {Olsen},\ and\ \citenamefont
  {Visscher}(2003)}]{fleig2003generalized}%
  \BibitemOpen
  \bibfield  {author} {\bibinfo {author} {\bibfnamefont {T.}~\bibnamefont
  {Fleig}}, \bibinfo {author} {\bibfnamefont {J.}~\bibnamefont {Olsen}}, \ and\
  \bibinfo {author} {\bibfnamefont {L.}~\bibnamefont {Visscher}},\ }\href@noop
  {} {\bibfield  {journal} {\bibinfo  {journal} {J. Chem. Phys}\ }\textbf
  {\bibinfo {volume} {119}},\ \bibinfo {pages} {2963} (\bibinfo {year}
  {2003})}\BibitemShut {NoStop}%
\bibitem [{\citenamefont {Visscher}, \citenamefont {Lee},\ and\ \citenamefont
  {Dyall}(1996)}]{visscher1996formulation}%
  \BibitemOpen
  \bibfield  {author} {\bibinfo {author} {\bibfnamefont {L.}~\bibnamefont
  {Visscher}}, \bibinfo {author} {\bibfnamefont {T.~J.}\ \bibnamefont {Lee}}, \
  and\ \bibinfo {author} {\bibfnamefont {K.~G.}\ \bibnamefont {Dyall}},\
  }\href@noop {} {\bibfield  {journal} {\bibinfo  {journal} {J. Chem. Phys.}\
  }\textbf {\bibinfo {volume} {105}},\ \bibinfo {pages} {8769} (\bibinfo {year}
  {1996})}\BibitemShut {NoStop}%
\bibitem [{\citenamefont {Shee}, \citenamefont {Visscher},\ and\ \citenamefont
  {Saue}(2016)}]{shee_analytic_2016}%
  \BibitemOpen
  \bibfield  {author} {\bibinfo {author} {\bibfnamefont {A.}~\bibnamefont
  {Shee}}, \bibinfo {author} {\bibfnamefont {L.}~\bibnamefont {Visscher}}, \
  and\ \bibinfo {author} {\bibfnamefont {T.}~\bibnamefont {Saue}},\ }\href
  {\doibase 10.1063/1.4966643} {\bibfield  {journal} {\bibinfo  {journal} {J.
  Chem. Phys.}\ }\textbf {\bibinfo {volume} {145}},\ \bibinfo {pages} {184107}
  (\bibinfo {year} {2016})}\BibitemShut {NoStop}%
\bibitem [{\citenamefont {Shee}\ \emph {et~al.}(2018)\citenamefont {Shee},
  \citenamefont {Saue}, \citenamefont {Visscher},\ and\ \citenamefont {Severo
  Pereira~Gomes}}]{shee_equation--motion_2018}%
  \BibitemOpen
  \bibfield  {author} {\bibinfo {author} {\bibfnamefont {A.}~\bibnamefont
  {Shee}}, \bibinfo {author} {\bibfnamefont {T.}~\bibnamefont {Saue}}, \bibinfo
  {author} {\bibfnamefont {L.}~\bibnamefont {Visscher}}, \ and\ \bibinfo
  {author} {\bibfnamefont {A.}~\bibnamefont {Severo Pereira~Gomes}},\ }\href
  {\doibase 10.1063/1.5053846} {\bibfield  {journal} {\bibinfo  {journal} {J.
  Chem. Phys.}\ }\textbf {\bibinfo {volume} {149}},\ \bibinfo {pages} {174113}
  (\bibinfo {year} {2018})}\BibitemShut {NoStop}%
\bibitem [{\citenamefont
  {Pernpointner}(2014)}]{pernpointner_relativistic_2014}%
  \BibitemOpen
  \bibfield  {author} {\bibinfo {author} {\bibfnamefont {M.}~\bibnamefont
  {Pernpointner}},\ }\href {\doibase 10.1063/1.4865964} {\bibfield  {journal}
  {\bibinfo  {journal} {J. Chem. Phys.}\ }\textbf {\bibinfo {volume} {140}},\
  \bibinfo {pages} {084108} (\bibinfo {year} {2014})}\BibitemShut {NoStop}%
\bibitem [{\citenamefont {Pernpointner}, \citenamefont {Visscher},\ and\
  \citenamefont {Trofimov}(2018)}]{pernpointner_four-component_2018}%
  \BibitemOpen
  \bibfield  {author} {\bibinfo {author} {\bibfnamefont {M.}~\bibnamefont
  {Pernpointner}}, \bibinfo {author} {\bibfnamefont {L.}~\bibnamefont
  {Visscher}}, \ and\ \bibinfo {author} {\bibfnamefont {A.~B.}\ \bibnamefont
  {Trofimov}},\ }\href {\doibase 10.1021/acs.jctc.7b01056} {\bibfield
  {journal} {\bibinfo  {journal} {J. Chem. Theory Comput.}\ }\textbf {\bibinfo
  {volume} {14}},\ \bibinfo {pages} {1510} (\bibinfo {year}
  {2018})}\BibitemShut {NoStop}%
\bibitem [{\citenamefont {Yuan}\ and\ \citenamefont
  {Gomes}()}]{dataset-lasercooling}%
  \BibitemOpen
  \bibfield  {author} {\bibinfo {author} {\bibfnamefont {X.}~\bibnamefont
  {Yuan}}\ and\ \bibinfo {author} {\bibfnamefont {A.~S.~P.}\ \bibnamefont
  {Gomes}},\ }\href {\doibase 10.5281/zenodo.6376250} {\enquote {\bibinfo
  {title} {Dataset: Reassessing the potential of tlcl for laser cooling
  experiments via four-component correlated electronic structure
  calculations},}\ }\bibinfo {howpublished} {\url
  {https://doi.org/10.5281/zenodo.6376250}}\BibitemShut {NoStop}%
\bibitem [{\citenamefont {Liu}\ \emph {et~al.}(2020)\citenamefont {Liu},
  \citenamefont {Yuan}, \citenamefont {Xiao}, \citenamefont {Xu},\ and\
  \citenamefont {Yan}}]{liu2020electronic}%
  \BibitemOpen
  \bibfield  {author} {\bibinfo {author} {\bibfnamefont {Y.}~\bibnamefont
  {Liu}}, \bibinfo {author} {\bibfnamefont {X.}~\bibnamefont {Yuan}}, \bibinfo
  {author} {\bibfnamefont {L.}~\bibnamefont {Xiao}}, \bibinfo {author}
  {\bibfnamefont {H.}~\bibnamefont {Xu}}, \ and\ \bibinfo {author}
  {\bibfnamefont {B.}~\bibnamefont {Yan}},\ }\href@noop {} {\bibfield
  {journal} {\bibinfo  {journal} {J. Quant. Spectrosc. Ra}\ }\textbf {\bibinfo
  {volume} {243}},\ \bibinfo {pages} {106817} (\bibinfo {year}
  {2020})}\BibitemShut {NoStop}%
\bibitem [{\citenamefont {Tiemann}(1988)}]{tiemann1988spectroscopic}%
  \BibitemOpen
  \bibfield  {author} {\bibinfo {author} {\bibfnamefont {E.}~\bibnamefont
  {Tiemann}},\ }\href@noop {} {\bibfield  {journal} {\bibinfo  {journal} {Mol.
  Phys}\ }\textbf {\bibinfo {volume} {65}},\ \bibinfo {pages} {359} (\bibinfo
  {year} {1988})}\BibitemShut {NoStop}%
\bibitem [{\citenamefont {Nguyen}\ \emph {et~al.}(2011)\citenamefont {Nguyen},
  \citenamefont {Viteri}, \citenamefont {Hohenstein}, \citenamefont {Sherrill},
  \citenamefont {Brown},\ and\ \citenamefont {Odom}}]{nguyen2011challenges}%
  \BibitemOpen
  \bibfield  {author} {\bibinfo {author} {\bibfnamefont {J.~H.~V.}\
  \bibnamefont {Nguyen}}, \bibinfo {author} {\bibfnamefont {C.~R.}\
  \bibnamefont {Viteri}}, \bibinfo {author} {\bibfnamefont {E.~G.}\
  \bibnamefont {Hohenstein}}, \bibinfo {author} {\bibfnamefont {C.~D.}\
  \bibnamefont {Sherrill}}, \bibinfo {author} {\bibfnamefont {K.~R.}\
  \bibnamefont {Brown}}, \ and\ \bibinfo {author} {\bibfnamefont
  {B.}~\bibnamefont {Odom}},\ }\href@noop {} {\bibfield  {journal} {\bibinfo
  {journal} {New. J. Phys}\ }\textbf {\bibinfo {volume} {13}},\ \bibinfo
  {pages} {063023} (\bibinfo {year} {2011})}\BibitemShut {NoStop}%
\bibitem [{\citenamefont {Fu}\ \emph {et~al.}(2017)\citenamefont {Fu},
  \citenamefont {Ma}, \citenamefont {Cao},\ and\ \citenamefont
  {Bian}}]{fu2017laser}%
  \BibitemOpen
  \bibfield  {author} {\bibinfo {author} {\bibfnamefont {M.}~\bibnamefont
  {Fu}}, \bibinfo {author} {\bibfnamefont {H.}~\bibnamefont {Ma}}, \bibinfo
  {author} {\bibfnamefont {J.}~\bibnamefont {Cao}}, \ and\ \bibinfo {author}
  {\bibfnamefont {W.}~\bibnamefont {Bian}},\ }\href@noop {} {\bibfield
  {journal} {\bibinfo  {journal} {J. Chem. Phys.}\ }\textbf {\bibinfo {volume}
  {146}},\ \bibinfo {pages} {134309} (\bibinfo {year} {2017})}\BibitemShut
  {NoStop}%
\bibitem [{\citenamefont {Norcia}\ \emph {et~al.}(2018)\citenamefont {Norcia},
  \citenamefont {Cline}, \citenamefont {Bartolotta}, \citenamefont {Holland},\
  and\ \citenamefont {Thompson}}]{norcia2018narrow}%
  \BibitemOpen
  \bibfield  {author} {\bibinfo {author} {\bibfnamefont {M.~A.}\ \bibnamefont
  {Norcia}}, \bibinfo {author} {\bibfnamefont {J.~R.}\ \bibnamefont {Cline}},
  \bibinfo {author} {\bibfnamefont {J.~P.}\ \bibnamefont {Bartolotta}},
  \bibinfo {author} {\bibfnamefont {M.~J.}\ \bibnamefont {Holland}}, \ and\
  \bibinfo {author} {\bibfnamefont {J.~K.}\ \bibnamefont {Thompson}},\
  }\href@noop {} {\bibfield  {journal} {\bibinfo  {journal} {New. J. Phys}\
  }\textbf {\bibinfo {volume} {20}},\ \bibinfo {pages} {023021} (\bibinfo
  {year} {2018})}\BibitemShut {NoStop}%
\end{thebibliography}%

\end{document}